%% file: paper.tex
\documentclass[12pt]{article}
\usepackage{amssymb,amsmath}
\usepackage{openbib}
\usepackage{amsthm}
\usepackage{graphicx}
\theoremstyle{plain}
\newtheorem{lem}{Lemma}
\newtheorem{prop}{Proposition}

\newtheorem{theor}{Theorem}

\newtheorem{defn}{Definition}

\def\dst{\displaystyle}

\def\cf{{\it cf. }}
\def\eg{{e.g.\ }}
\def\ie{i.e.\ }

\def\wig{_{{\rm\scriptstyle w}}}
\def\sf{{\textstyle{\frac 1 2}}}
\def\R{{\mathbb{R}}}      
\def\C{{\mathbb{C}}}      
\def\hatw{\widehat}       
\def\wt{\widetilde}       
\def\grad{{\nabla}}       
\def\H{\mathcal{H}}       
\def\S{\mathcal{S}}

\def\M{\mathcal{M}}
\def\N{\mathcal{N}}
\def\RR{\mathcal{R}}
\def\K{\mathcal{K}}
\def\A{\mathcal{A}}

\def\I{\mathcal{I}}
\def\R{\mathbb{R}}
\def\s{{\it s}}
\def\p{\partial}
\def\({{\Bigl(}}          
\def\){{\Bigr)}}          

\def\X{X}
\def\Y{Y}
\def\st{\stackrel}
\def\itn#1{{\rm{(#1)}}}    

\def\theequation{\arabic{section}.\arabic{equation}}
\begin{document}
\input{title}
\input{sec1}
\input{sec2}
\input{sec3}
\input{sec4}
\input{sec5}
\renewcommand{\theequation}{\Alph{section}.\arabic{equation}}
\appendix

\input{appendix1}

\end{document}

%% file: title.tex
\begin{titlepage}

\font\title=cmbx12 \centerline{{\title Heisenberg Evolution WKB and Symplectic Area Phases}}
\vspace{15 mm}

\centerline{T. A. Osborn }
\vskip 8pt \centerline{{\sl Department of Physics and Astronomy}}
\centerline{{\sl University of Manitoba}}
\centerline{{\sl Winnipeg, MB, Canada, R3T 2N2 }}
\vspace{10mm}

\centerline{M. F. Kondratieva }
\vskip 8pt \centerline{{\sl Department of Mathematics and Statistics}}
\centerline{{\sl University of Minnesota}}
\centerline{{\sl Duluth, MN, USA 55812}}

\vspace{10 mm}

\begin{abstract}

The Schr\"odinger and Heisenberg  evolution operators are represented in phase space $T^*\R^n$ by
their Weyl symbols. Their semiclassical approximations are constructed in the short and long time
regimes. For both evolution problems, the WKB representation is purely geometrical: the
amplitudes are functions of a Poisson bracket and the phase is the symplectic area of a region in
$T^*\R^n$ bounded by trajectories and chords. A unified approach to the Schr\"odinger and
Heisenberg semiclassical evolutions is developed by introducing an extended phase space
$\chi_2\equiv  T^*(T^*\R^n)$. In this setting Maslov's pseudodifferential operator version of WKB
analysis applies and represents these two problems via a common higher dimensional Schr\"odinger
evolution, but with different extended Hamiltonians. The evolution of a Lagrangian manifold in
$\chi_2$, defined by initial data, controls the phase, amplitude and caustic behavior. The
symplectic area phases arise as a solution of a boundary condition problem in $\chi_2$. Various
applications and examples are considered. Physically important observables generally have symbols
that are free of rapidly oscillating phases. The semiclassical Heisenberg evolution in this
context has been traditionally realized as an $\hbar$ power series expansion whose leading term
is classical transport. The extended Heisenberg Hamiltonian has a reflection symmetry that
ensures this behavior. If the WKB initial phase is zero, it remains so for all time, and
semiclassical dynamics reduces to classical flow plus finite order $\hbar$ corrections.

\vfill

\end{abstract}
\end{titlepage}

%% file: sec1.tex

\section{Introduction}

The version of quantum mechanics most suitable for semiclassical dynamics is Moyal's \cite{Moy49}
representation.  This classical like statement of quantum mechanics is achieved by the
Wigner--Weyl \cite{Wig32,Wey27} correspondence, which uniquely associates Hilbert space operators
with Weyl symbols (functions) on phase space. This correspondence transforms the non-commutative
product of operators into a non-commutative $*$ product of symbols.  Expectation values are
realized as integrals over phase space.

For systems whose coordinate manifold is flat, the Wigner--Weyl isomorphism is most compactly
described \cite{Gro76,Roy77,GH78} in terms of the quantizer. Let $ T(x) = \exp[i(p\cdot\hat q -
q\cdot\hat p)/\hbar]$ be the Heisenberg translation operator. Here $x = (q,p)$ are the position,
momentum coordinates of the linear phase space $\chi_1\equiv T^*\R^n_q$, and $\hat x = (\hat
q,\hat p)$ are the corresponding quantum counterparts. The parity operator on Hilbert space is
$P\psi(q)= \psi(-q)\,, \psi\in L^2(\R^n_q)$. The product of these two unitary operators defines
the quantizer, $\Delta(x)\equiv 2^n T(2x)P$. In terms of $\Delta(x)$, the one-to-one Weyl
symbol-operator pairing is
\begin{equation*}\label{eq1.0}
    \hatw A = h^{-n} \int_{\chi_1} dx A(x) \Delta(x)\,,\qquad A(x) = [\hatw A\,]\wig(x)
    = {\rm{Tr}}\, \hatw A\,\Delta(x)\,,
\end{equation*}
where $h= (2\pi \hbar)\,$. The first statement above is Weyl quantization; while the second is
de-quantization. Whenever the operator is a density matrix $\hat\rho$, the quantity
$h^{-n}\rho(x)\,$ is the Wigner distribution. The above map, from $\hatw A$ to $A(x)$, gives the
same result as the Wigner transform (\ref{eqnA3}) of the kernel $\langle q|\hatw A |q' \rangle$.

The star product is defined by the requirement that $[\hatw A \hatw B\,]\wig = A*B$ \cf  Appendix
A. For small $\hbar$, the $*$ multiplication is approximately the commutative product of symbols,
\begin{equation*} A*B(x) = A(x)B(x) + {{i\hbar}/2}\{A,B\}(x) + O(\hbar^2)\, .
\end{equation*}
The Poisson bracket term measures, to leading $\hbar$ order, the non-commutative character of the
star product.

The two basic statements of quantum dynamics are the Schr\"odinger and Heisenberg pictures. On
the space, $L^2(\R^n_q)$, each self-adjoint Hamiltonian $\hatw H$ generates a unitary
Schr\"odinger evolution, $\hatw U(t)=\exp[-it\hatw H/\hbar]$.  An initial density matrix $\hat
\rho_0$ has Heisenberg evolution, $\hat \rho(t) =\hatw U(t)\hat \rho_0 \hatw U(t)^\dag$. Let $H$,
$U(t)$ and $\rho(t)$ be the Weyl symbols of $\hatw H$, $\hatw U(t)$ and $\hat \rho(t)$,
respectively. In this setting the equations of motion are
 \begin{eqnarray} \label{eq1.4} i\hbar\!\!\!\!&{\p U(t,x)}/{\p t}\!\!  &= H(x)\! * U(t,x) \, ,
\\ \label{eq1.5}
 &{\p\rho(t,x)}/{\p t}\!\!  &=\{H, \rho(t)\}_M (x)    \,,
\end{eqnarray}
with initial conditions $U(0,x) = 1$,  and $\rho(0,x) = \rho_0(x)$.  In (\ref{eq1.5}), the bracket
operation is defined by
\begin{equation*}\label{eq1.6}
    \{A,B\}_M \equiv \frac 1 {i\hbar} [\hatw A,\hatw B\,]\wig = \frac 1 {i\hbar} (A*B -B*A)   \, .
\end{equation*}
This is the Moyal bracket. Like the quantum commutator, it is bilinear, skew and obeys the Jacobi
identity.

It is evident that the Weyl--Heisenberg symbol evolution problem (\ref{eq1.5})  is formally as
close as it can be to classical dynamics. Since $\{A,B\}_M = \{A,B\} + O(\hbar^2)\,$, Bohr's
correspondence principle is realized in a transparent form. The time dependent expectation value
of an observable $\hatw A$ is conveniently given by the phase space integral
\begin{equation}
\langle \hatw A \rangle_t \equiv {\rm Tr}\,\hatw A\,\hat \rho(t)= h^{-n}\int_{\chi_1}
dx\,A(x)\rho(t,x)\,.
\end{equation}

The main goal of this paper is to obtain both short and long time WKB approximate solutions for
the Weyl symbol evolution problems (\ref{eq1.4}) and (\ref{eq1.5}).  In Theorems 1 and 2, we
prove that the phases entering these semi-classical representations are invariant geometrical
quantities --- namely the symplectic areas defined by certain closed loops in phase space.
Furthermore, the amplitude functions are given by determinants of a Poisson bracket.

We employ a common method to solve both problems (\ref{eq1.4}) and (\ref{eq1.5}). Standard Weyl
symbol calculus identities show that $H*$ and $\{H,\cdot\}_M$ are pseudodifferential operators,
in other words they are functions of $\grad_x$ and multiplication by $x$. If one introduces an
extended phase space $\chi_2\equiv T^*(T^*\R^n_q)$, then Maslov's \cite{MF81} WKB analysis
applies.  In particular, the semiclassical approximations for the evolutions $U(t,x)$ and
$\rho(t,x)$ are characterized by a Lagrangian manifold, and the time evolution of this manifold
is determined by an extended Hamiltonian in $\chi_2$. In order to distinguish these two phase
spaces we call $\chi_1$ the primary phase space (PPS), and $\chi_2$ the secondary phase space
(SPS). Projecting the $\chi_2$ analysis back onto the primary phase space provides explicit WKB
representations of $U(t,x)$ and $\rho(t,x)$.

The results in the literature nearest to this paper are found in three seminal works: by Karasev
and Nazaikinskii \cite{KN78}, Berry \cite{Ber77}, and Marinov \cite{Mar79}. In the first of
these, an analytical approach to a generalized semiclassical expansion based on a SPS with its
associated left-right representation of $*$-product and Hamilton-Jacobi equation was developed.
In the second pair of works, the geometrical (symplectic phase) interpretation of the Wigner
eigenfunction and of the dynamical WKB approximation was achieved. Specifically, in the small
time sector, where the time evolving $\chi_2$ Lagrangian manifold remains single sheeted, Marinov
found the WKB amplitude and phase for $U(t,x)$.  In addition to providing an entirely different
proof, our treatment extends these known results in several ways. First, we determine the long
time version of the WKB approximation where the $\chi_2$ Lagrangian manifold may be multi-sheeted
with several points corresponding to a given $x\in\chi_1$. Second, by incorporating Maslov's
results one maintains control of the error estimates. Third, we solve a generalized version of
(\ref{eq1.4}) with rapidly oscillating initial data. This latter generalization arises for the
evolution $\hatw U(t+t_0),\, t_0>0$.  At $t=0$ the initial data generically will have the rapidly
oscillating form, $N_0(x)e^{ i\Phi_0(x)/\hbar}$.

In spite of the importance of problem (\ref{eq1.5}), it has received insufficient study. The
version of the Heisenberg problem that has been investigated in detail
\cite{Ant77,Pro83,DR87,BS91} occurs for the case where the observable $\hatw A$ is
semiclassically admissible, namely its Weyl symbol is $\hbar$ dependent and allows a finite order
small $\hbar$ asymptotic expansion \cf (\ref{eq3.7a}). In \cite{OM95} a connected graph method
was constructed to obtain the coefficients of the $\hbar$ asymptotic  expansion of $[\hatw
U(t)^\dag \hatw A \hatw U(t)]\wig(x)$. Based on this latter formalism, numerical calculations
have shown \cite{McQ98} that the $\hbar^2$-order semiclassical expansion accurately describes
noble gas atom-atom scattering. However, semiclassically admissible symbols are not suitable for
representing \cite{Hel76} density matrices. The requirement that one has a pure state, $\rho_0 *
\rho_0 = \rho_0$, means that $\rho_0(x)$ is an $\hbar \rightarrow 0$ rapidly oscillating function
\cf (\ref{eq2.19}).

The paper has the following organization.  Section 2 presents the extended Schr\"odinger equation
in the SPS setting and shows how the problems (\ref{eq1.4}) and (\ref{eq1.5}) are particular
examples of this extended evolution.  Also, the geometric, commutative and coordinate
relationships between the PPS and SPS are reviewed. The trajectories entering the WKB expansion
are determined by the solutions of a boundary condition problem.  In Section 3, a detailed
analysis of the BC problem and the manner in which it constructs phase space loops is carried
out. The semiclassical expansions for $U(t,x)$ and $\rho(t,x)$ are consolidated in Theorems 1 and
2. Finally, in Section 4 the group generated phase addition rules are obtained; the mutual
compatibility of the Schr\"odinger and Heisenberg semiclassical evolution is established; the
exact solutions for quadratic Hamiltonian systems are compared; and, the $\hbar \rightarrow 0$
asymptotics of $[\hatw U(t) \hatw A \hatw U(t)^\dag]\wig(x)$ for semiclassically admissible
operators $\hatw A$ is realized as a special case of Theorem 2. There are three appendices
containing specialized aspects of Weyl symbol calculus, pseudodifferential operators, Jacobi
fields, and a Poincar\'e--Cartan identity.

%% file: sec2.tex
\section{Secondary Phase Space Dynamics}
\setcounter{equation}{0}

In the first part of this section we show how to interpret the $H*$ structure in the equations of
motion as the action of a pseudodifferential operator ($\Psi$DO).  The normal ordered symbols of
these operators are then functions on the secondary phase space.  For the Schr\"odinger and
Heisenberg problem we show how the classical flow in $\chi_2$  is composed in terms of the
standard Hamiltonian mechanics on $\chi_1$. The necessity of representing the initial value form
of the density matrix as a rapidly oscillating $\hbar$ function is confirmed by constructing the
semiclassical Weyl symbol for a pure state.

Introduce the SPS canonical operators $\X,\Y$ acting on functions $f(x)$: $\X f(x) = xf(x)$ and
$\Y f(x) = -i\hbar \grad_x f(x)$. Let $\H(\st{2}{X},\st{1}{Y})$ be a normal ordered, symmetric
$\Psi$DO defined by the symbol $\H(x,y)$. On the coordinate domain $\R^{2n}_x$, the extended
Schr\"odinger evolution problem is
\begin{equation} \label{eq2.0}
i\hbar{\p \Psi(t,x)}/{\p t}= \H(\st{2}\X,\st{1}\Y)\Psi(t,x) \,.
\end{equation}
The two problems (\ref{eq1.4}) and (\ref{eq1.5}) are particular cases of (\ref{eq2.0}). The
appropriate realization of $\H$ in each case is determined by employing the left-right $*$
product representation \cite{KN78,KM91} (see Appendix A). Setting $\H(x,y)=\H_1 \equiv H(x-\frac
12 Jy)$ selects the Schr\"odinger problem (\ref{eq1.4}), whereas the choice $\H_2 \equiv
H(x-\frac 12 Jy)-H(x+\frac 12 Jy)$ gives the Heisenberg problem (\ref{eq1.5}). The solution
$\Psi(t,x)$ is either $U(t,x)$ or $\rho(t,x)$ depending on the choice of $\H$. In SPS, the scalar
function $\H$ is a Hamiltonian which is the generator of $\chi_2$ classical flow.

Consistency of this extended Schr\"odinger problem requires that the normal ordered symbol
$H(x\pm\frac 12 Jy)$ define a symmetric operator on $L^2(\R_x^{2n})$.  That is does so is a
consequence of the argument structure $(x\pm \frac 12 Jy)$ which implies that \cf (\ref{ATR}) the
normal ordered and Weyl ordered symbols are the same function. Since the Weyl ordered symbol is
real, one has that the $\Psi$DO Hamiltonians in (\ref{eq2.0}) are formally self-adjoint.

    Consider the coordinate systems of $\chi_1$ and $\chi_2$ and their interconnections. Let $y$ be
the momentum variable in SPS so that a general point $m\in \chi_2$ has Cartesian coordinates
$z(m)=(x,y)$. The symplectic structure of $\chi_1$ and $\chi_2$ is determined by their respective
canonical 2-forms
\begin{eqnarray*}\label{eq2.4}
&\omega^{(1)} \equiv dp\wedge dq  =\frac 12 J_{\alpha\beta}\, dx_\beta\wedge dx_\alpha\,,\quad J
= \left[\begin{array}{cc} 0&I_n\\-I_n&0\end{array}\right]\,, \\ \label{eqn2.6} &\omega^{(2)}
\equiv dy\wedge dx  =\frac 12 \wt J_{jk}\, dz_k\wedge dz_j\,,
\end{eqnarray*}
where $I_n$ is the $n\times n$ identity matrix, and $\wt J$ is the symplectic matrix having the
block form of $J$ with $I_{2n}$ substituted for $I_n$. Given the 2-forms, $\omega^{(1)}$ and
$\omega^{(2)}$, the corresponding Poisson brackets are
\begin{equation*}\label{eq2.7} \{f,g\}_1(x) =\nabla f(x)\cdot J \nabla g(x)\,,\qquad
\{F,G\}_2(z) =\nabla F(z)\cdot \wt J \nabla G(z)\,.
\end{equation*}
The coordinate functions, $x=(q,p)$ and $z=(x,y)$, are canonical in $\chi_1$ and $\chi_2$, namely
$\{x_\alpha ,x_\beta\}_1 = J_{\alpha\beta}$ and $\{z_j, z_k\}_2 = \wt J_{jk}$. In order that
(\ref{eq2.0}) be a standard Schr\"odinger equation the coordinates $x=(q,p)$ must commute.  This
is the case since $\{x_\alpha , x_\beta\}_2 = 0$.

As one sees from functional forms of $\H$,  a second natural coordinate system for $\chi_2$ is
defined by the $(l,r)$ variables
\begin{equation}\label{eq2.8}
\begin{array}{l}
\dst l=x- \sf J y, \quad \dst r=x+{\textstyle{\frac 1 2}} J y\,, \quad \dst l,r\in \R^{2n}\,, \\[2ex]
    x={}\frac 1 2 (l+r)\,, \quad y= J (l-r)\,.
\end{array}
\end{equation}
 The link
between the coordinate systems $(x,y)$ and $(l,r)$, and their commutative structures is the
following.

\begin{lem}\label{lem1}
\noindent \itn{i} The SPS commutative properties of the projections $l,r :\R_z^{4n} \to
\R_x^{2n}$ are
\begin{equation*}
 \{l_\alpha,l_\beta\}_2=J_{\alpha\beta}\,, \quad \{r_\alpha,r_\beta\}_2=-J_{\alpha\beta}\,, \quad
\{l_\alpha,r_\beta\}_2=0\,.
\end{equation*}

\noindent \itn{ii} The $l$ variables are Poisson; the $r$ variables are anti-Poisson, namely
\begin{equation*} \{f\circ l,g\circ l\}_2=\{f,g\}_1\circ l, \quad \{f\circ r,g\circ r\}_2=-\{f,g\}_1\circ r.
\end{equation*}

\noindent \itn{iii} The $\chi_2 $ Poisson bracket and symplectic form have the $l\times r$
decomposition
\begin{eqnarray*}
 &\{F,G\}_2 = (\nabla_l F)_\alpha J_{\alpha\beta} (\nabla_l G)_\beta- (\nabla_r F)_\alpha
J_{\alpha\beta} (\nabla_r G)_\beta\,, \\
 &\wt J_{jk}\, dz_k\wedge dz_j =
J_{\alpha\beta}\, dl_\beta\wedge dl_\alpha- J_{\alpha\beta}\, dr_\beta\wedge dr_\alpha\,.
 \end{eqnarray*}
\end{lem}

The space structure for the left-right variables is $\chi_1\otimes\chi_1' \ni (l,r)$.  Here
$\chi_1'$ denotes the phase space defined by the 2-form $-\frac 12 J_{\alpha\beta}\,
dr_\beta\wedge dr_\alpha$.  Often the space $\chi_1\otimes\chi_1' $ is labelled  \cite{Litj90}
the double phase space. We use the name secondary phase space for cotangent bundle
$\chi_2=T^*(T^*R_q^n)$ in order to distinguish it from $\chi_1\otimes\chi_1'$. The left-right
map, (\ref{eq2.8})
is diffeomorphic.

The basic ingredients in any semiclassical approximation are the classical action phase and the
amplitude functions determined by classical transport.  Because the extended Hamiltonians are the
simple composite functions $H\circ l$ and $H\circ l -H\circ r$, it is possible to describe the
$\chi_2$ phase space motion as appropriately weighted sums of $\chi_1$ trajectories. Let
$g(\tau|x_0) = (q(\tau|x_0);p(\tau|x_0))$ denote the solution of the PPS Hamiltonian system
\begin{equation}\label{eq2.12}
    \dot g(\tau|x_0) = J\grad H(g(\tau|x_0))\,, \qquad g(0|x_0)=x_0\,.
\end{equation}
The SPS motion is labelled $G(\tau|z_0) = (x(\tau|z_0);y(\tau|z_0))$, where
\begin{equation}\label{eq2.13}
    \dot G(\tau|z_0) = \wt J\grad \H(G(\tau|z_0))\,, \qquad G(0|z_0)=z_0\,.
\end{equation}
The initial data $z_0$ for system (\ref{eq2.13}) also has the left-right representation $l_0 =
l(z_0)\,, r_0 = r(z_0)$.

\begin{lem}\label{lem2} The solution of the SPS Hamiltonian system, in terms of $\chi_1$ flows, takes the
form{\rm :} \noindent \itn{i} For the Schr\"odinger problem with $\H_1 = H\circ l$
\begin{equation}\label{eq2.14}
x(\tau|z_0) = \sf \big(g(\tau|l_0) + r_0\big)\,, \qquad y(\tau|z_0) = J\big(g(\tau|l_0)
-r_0\big)\,.
\end{equation}

\noindent \itn{ii} For the Heisenberg problem with $\H_2 = H\circ l - H\circ r$
\begin{equation}\label{eq2.15}
x(\tau|z_0) = \sf \big(g(\tau|l_0) + g(\tau|r_0)\big)\,, \qquad y(\tau|z_0) =
J\big(g(\tau|l_0) -g(\tau|r_0)\big)\,.
\end{equation}
\end{lem}

\noindent {\it Proof}. Consider part (i). Using the Poisson equations of motion, it is convenient
to state the SPS flow in the $l,r$ variables. Lemma \ref{lem1}(iii) implies
\begin{equation}\label{eq2.15b}
    \dot l_\alpha(\tau) = \{l_\alpha, H\circ l\}_2 =
    J_{\alpha\beta}\grad_\beta H (l(\tau))\,,\qquad \dot r_\alpha(\tau)
     = \{r_\alpha, H\circ l\}_2 = 0\,.
\end{equation}
So $l(\tau|l_0) = g(\tau|l_0)$ and the $r$-motion is a constant, $r(\tau|r_0) = r_0$. Using
(\ref{eq2.8}) recovers (\ref{eq2.14}).  A similar argument verifies part (ii). $\quad\square$

The identities (\ref{eq2.15b}) show that the left-right motions are decoupled. This decoupling
occurs because $\{l_\alpha(\tau),r_\beta(\tau)\}_2 = 0$. Further note that the system $\H_1=
H\circ l$ has $2n$ constants of motion; $r_\alpha(\tau)=const$ for $\alpha = 1\dots 2n$.
Nevertheless, $\H_1$ is not completely integrable since the functions $r_\alpha(\tau)$ are not in
involution, i.e. $\{r_\alpha(\tau),r_\beta(\tau)\}_2 = -J_{\alpha\beta}\neq 0$.

Lagrangian manifolds in SPS create canonical transformations in PPS. Let us clarify how this
is realized in terms the left right projections.  Suppose $\Lambda$ is a $\chi_2$ manifold
which is locally defined by the smooth phase $S:U\subseteq \R_x^{2n} \rightarrow \R$, \ie
$y=\grad S(x), x\in U$. This identity relates $l$ to $r$ by
\begin{equation}\label{eq2.15x}
r-l = J \grad S(\sf(l+r))\,.
\end{equation}
If $\det(I\pm \frac 12 JS'') \neq 0\,,\, x\in U$ then the implicit function theorem defines
$l=l(r)$ and $r=r(l)$. These transformations are canonical.  This follows from
\begin{equation*}
\frac{dl}{dr} = \frac {(I-\frac 12 JS'')}{(I+\frac 12 JS'')}\,.
\end{equation*}
The right side above is the Cayley transform of the symmetric $S''$ and so defines a
symplectic matrix. A systematic study of generating functions of this type is found in \cite{AH80}

    An important aspect of WKB analysis is the $\hbar$ singularity structure of the Cauchy initial
data. As noted in the Introduction, a pure state density matrix must be a rapidly oscillating
function. Let us clarify this situation with an example. Suppose $\psi_0(q) = n(q)\exp[
i\s(q)/{\hbar}]$ is a unit normalized $L^2(\R^n_q)$ wave function,  which defines a rank one
density matrix, $\hat \rho_0 = |\psi_0\rangle \langle \psi_0|$. The functions $n,\s$ are real.
The Weyl symbol of $\hat \rho_0$ is given by the Wigner transform (\ref{eqnA3})
\begin{eqnarray}\label{eq2.16}
&\rho_0(x;\hbar)  &= \int_{\R^n} dv\,n(q+ \sf v)n(q - \sf v)\exp {{\frac i{\hbar}}
\phi(v,x)}\,,\\ \label{eq2.17} &\phi(v,x) &= \s(q+ \sf v) - \s(q- \sf v) - p\cdot v\,.
\end{eqnarray}

The $\hbar \rightarrow 0$ asymptotic form of $\rho_0$ may be calculated by a  stationary phase
approximation \cite{MF81}. Let $\phi'(v,x) = \grad_v \phi(v,x)$ and $\phi''(v,x)= \grad_v \grad_v
\phi(v,x)$. The critical points of $\phi$ satisfy
\begin{equation}\label{eq2.18}
    2p = \grad \s(q-\sf v) + \grad\s(q+\sf v)\,.
\end{equation}
Suppose $(v_0,x_0)$ is a solution set for this equation where $\det \phi''(v_0,x_0) \neq 0$, then
the implicit function theorem defines a function $v=v(q,p)=v(x)$ obeying (\ref{eq2.18}). Since
$\phi'(v,x)$ is even in $v$, the roots occur in pairs. If $v=v(x)$ is a solution, then so is $v=
-v(x)$.  Incorporate this pairing behavior via the notation: $q_{\pm} =q \pm v(x)/2$, $p_{\pm} =
\grad \s(q_{\pm})$, $x_{\pm} = (q_{\pm},p_{\pm})$.  In the non-singular region where $\det
\phi''(v(x),x) \neq 0$, the Wigner transform above has asymptotic form
\begin{equation} \label{eq2.19}
\rho_0(x;\hbar) \approx h^{\frac n2}|\det\, \phi''(v(x),x)|^{-1/2} n(q_+)n(q_-)2\cos \Big\{\frac
{1}{\hbar}S_0(x) +\frac \pi{4} {\rm{sgn}} \, \phi''(v(x),x)\Big\}\,.
\end{equation}
The phase $S_0(x) = \phi(v(x),x)$ and sgn denotes the signature of a symmetric matrix. If $x$ is
in the neighborhood where (\ref{eq2.18}) has no solutions, then $\rho_0(x)$ is
$O(\hbar^{\infty})$ small.

As Berry first noticed \cite{Ber77} (in the $n=1$ version of this problem), the phase $S_0$ is a
symplectic area. To see this in the present context, define the $\chi_1$ Lagrangian manifold
$\lambda \equiv \{x\in \chi_1 | p = \grad \s(q)\}$. Construct two paths between the end points
$x_-$ and $x_+$. Let the first path, $C(x_-,x_+)$, be a straight line from $x_-$ to $x_+$. As the
second path, $\gamma_{\lambda}(x_+,x_-)$, choose any curve from $x_+$ to $x_-$ lying on the
surface $\lambda$. Then one finds
\begin{equation} \label{eq2.20}
    S_0(x) = \oint_{C(x_-,x_+)+\gamma_{\lambda}(x_+,x_-)} p\cdot dq\,.
\end{equation}
The geometry of the loop in (\ref{eq2.20}) is the well known chord mid-point construct
\cite{Mar79,Ber77,Ber89,OH81,OdA98}. The end points $x_{\pm}$ of the directed chord $C(x_-,x_+)$
lie on $\lambda$, and $x$ is the mid-point of this chord. By Stokes theorem, $S_0(x)$ is also the
area of any membrane having boundary $C(x_-,x_+)+\gamma_{\lambda}(x_+,x_-)$.

%% file: sec3.tex
\section{Symplectic Areas and  WKB Phases}
\setcounter{equation}{0}

    This section shows how the secondary phase space WKB approximations for $U(t,x)$ or $\rho(t,x)$ can
be reformulated in terms of primary phase space flows.  This reduction process transforms all the
$\chi_2$ WKB phases into symplectic areas defined by closed loops in $\chi_1$.  A key element of
this analysis is the solvability of the appropriate two point boundary condition (BC) problem. We
pose the BC problem in terms of Lagrangian manifolds and construct both short and long time
solutions.

To begin, we summarize Maslov's WKB expansion for the extended Schr\"odinger problem
(\ref{eq2.0}). The $t=0$ state for system (\ref{eq2.0}) is assumed to have the generic form,
$\Psi_0(x) = \alpha_0(x) e^{i\beta_0(x)/\hbar}$. The real functions $\alpha_0, \beta_0$ have
support on the domain $D_0\subseteq \R_x^{2n}$. The phase $\beta_0$ defines a $\chi_2$ Lagrangian
manifold $\Lambda_0$, via $y(x)=\grad\beta_0(x), x\in D_0$. Let $\Pi_1(x,y) = x$ be the
projection onto $\R_x^{2n}$.  It is assumed that projection $\Pi_1:\Lambda_0\rightarrow D_0$ is a
diffeomorphism.

To proceed three assumptions are required: (a) The function $\H \in T_{+}^m(\R_z^{4n}), m\geq 2$.
\cf Appendix A. (b) For arbitrary finite time intervals $[-T,T]$, the SPS Hamiltonian system
(\ref{eq2.13}) has unique smooth solutions, i.e. $G(t|x,y)\in C^{\infty}([-T,T],\R_z^{4n})$. (c)
Let $x$ be a non-focal point (\cf Definition 3) with respect to the $\Pi_1$ projection of the
manifold $\Lambda_t = G(t)\Lambda_0$.  Assume there are a finite number of points $x_0^j \in
D_0$, $j=1,\dots,N$ such that $\Pi_1 G(t|x_0^j,\grad\beta_0(x_0^j))= x$.

Each of these assumptions is required for an evident reason: (a) ensures that
$\H(\st{2}\X,\st{1}\Y)$ is a well defined $\Psi$DO, (b) prohibits finite time runaway
trajectories, and (c) assumes the existence of one or more solutions to the BC problem: which
trajectories starting from $\Lambda_0$ have $x$ as their final coordinate position? Locally
varying $(t,x)$ in (c) determines the initial $x_0$ as a function of $(t,x)$, namely
$x_0^j=x_0^j(t,x)$.

Under the hypotheses (a--c) the $\hbar \rightarrow 0$ asymptotic solution of the Cauchy problem
(\ref{eq2.0}), with initial data $\Psi_0$, is
\begin{eqnarray}\label{eq3.1}&\Psi(t,x)\,\, = &\Psi^{sc}(t,x)[1 + O(\hbar)\,]\,,\\ \label{eq3.1b}
\vspace{1 mm}
    &\Psi^{sc}(t,x) = &\sum_{j=1}^N \frac {\phi_j(t,x)}{\surd|J_j(t,x)|} \exp\Big\{\frac i{\hbar}
    \beta_j(t,x) -\frac {i\pi}2 m_j(t,x)\Big\}\,, \qquad t\in [-T,T]\,.
\end{eqnarray}
Here $m_j(t,x)$ is the Maslov index for trajectory $\{G(\tau|z_0^j):\tau\in [0,t]\}$, where $z_0^j
=(x_0^j(t,x),\grad\beta_0(x_0^j(t,x)))\in \Lambda_0$. The phase and amplitude functions are
\begin{eqnarray}\label{eq3.2a}
&\beta_j(t,x) &= \beta_0(x_0^j) + \int_0^t \left[ y(\tau|z_0^j)\cdot \dot x(\tau|z_0^j) -
\H(G(\tau|z_0^j))\right] d\tau \,, \\ \label{eq3.2b} &J_j(t,x) &= \det \bigg ( \frac
{dx(t|x_0^j,\grad\beta_0(x_0^j))}{dx_0^j} \bigg)\bigg |_{x_0^j= x_0^j(t,x)} \,,\\ \label{eq3.2c}
&\phi_j(t,x) &= \left( \exp {\int_0^t tr \grad_x \grad_y \H(G(\tau|z_0^j)) d\tau}\right)
\alpha_0(x_0^j)\,.
\end{eqnarray}

The family of operators $\H(\st{2}\X,\st{1}\Y)$ consistent with the class $T_{+}^m(\R_z^{4n})$ is
large. The growth index, $m$, must be 2 or greater so as to include kinetic energy in $H$. The
operators $\H(\st{2}\X,\st{1}\Y)$ need not have a polynomial dependence in momentum $\Y$, nor do
they need to be partial differential operators. This generality is essential to the SPS method we
employ: normally, the $\Psi$DO defined by $\H_1$ and $\H_2$ will not be a finite order partial
differential operator.  The phases $\beta_j(t,x)$ have a geometric meaning. Define the $\H$ flow
transported Lagrangian manifold, $\Lambda_t \equiv G(t)\Lambda_0$. Then the surfaces $y_j(t,x) =
\grad \beta_j(t,x)$ describe the different sheets of $\Lambda_t$. Representation
(\ref{eq3.1}--\ref{eq3.2c}) is Theorem 10.5 of \cite{MF81}.

In phase space quantum mechanics, the range of physical systems one may describe is controlled by
the allowed functional form of the Weyl symbol Hamiltonian. For reasons of notational simplicity
we presume $H$ is static. Including time dependent Hamiltonians, as a subsequent generalization,
is a straightforward matter. The WKB summary above indicates that two restrictions on $H$ are
necessary.

{\bf Assumption A}1.  The Hamiltonian operator $\hatw H$ is semiclassically admissible,
specifically $H \in T^m_{+}(\R_x^{2n})\,,  m\geq 2$. \quad {\bf A}2. On finite time intervals
$[-T,T]$, $H$ generates unique classical flow, $g(t|x)\in C^{\infty}([-T,T],\R_x^{2n})$.

\smallskip

Assumption {\bf A} is common to all the Propositions and Theorems of this section and will
not be explicitly cited in them.

Since $l$ and $r$ are linear functions of $z$, Assumption {\bf A}1 implies that  the composite
functions $\H_1 =H\circ l$ and $\H_2 = H\circ l - H\circ r$ are in $T_+^m(\R_z^{4n}), m\geq 2$.
Likewise, Lemma 2 shows that Assumption {\bf A}2 guarantees that $G(t|x,y)\in
C^{\infty}([-T,T],\R_z^{4n})$ for both $\H_1$ and $\H_2$ flow.  The requirement that $H \in
T^m_{+}(\R_x^{2n})$ (\cf Appendix A) means that $H=H(x;\hbar)$ has the $\hbar\rightarrow 0$
uniform asymptotic expansion
\begin{equation}\label{eq3.7a}
H(x;\hbar)= H_0(x) + \sum_{j=1}^J \hbar^j H_j(x) + O(\hbar^{J+1})\,.
\end{equation}
For many physical systems $H(x;\hbar)=H_0(x)$. For example, this occurs for a mass $m$ charged
particle moving in an external electromagnetic field.   The higher order terms $H_j(x)$ have no
effect on the leading order WKB approximation. So from here on we will identify $H$ with $H_0$
and omit the 0 subscript.

\subsection{The Boundary Condition Problem}

The assumption {\bf A} says nothing about the solvability of the BC problem. Rather than dealing
with this by assumption, as was done in the representation (\ref{eq3.1}--\ref{eq3.2c}), we will
find existence proofs and explicit formulas for the BC solutions.

    We discuss the BC conditions for the evolutions $U(t,x)$ and $\rho(t,x)$ in tandem. For the
$U(t,x)$ evolution, take the initial state of the system to be
\begin{equation}\label{eq3.8}
U_0(x)= N_0(x)e^{i \Phi_0(x)/\hbar}\,, \qquad x\in D_0\subseteq \R^{2n}\,
\end{equation}
where the domain, $D_0$, is a simply connected open set.  The phase $\Phi_0$ defines the single
sheeted Lagrangian manifold $\Lambda_0=\{z\in\chi_2|y(x)=\grad \Phi_0(x)\,, x\in D_0\}$. The
initial state for density matrix evolution $\rho(t,x)$ is presumed to be
 \begin{equation}\label{eq3.5}
\rho_0(x)= \alpha_0(x)e^{i S_0(x)/\hbar}\,, \qquad x\in D_0\subseteq \R^{2n}\,.
\end{equation} Again, $\Lambda_0$ will denote the simply connected manifold generated by $S_0$.
The phases $\Phi_0$ and $S_0$ are both assumed to be $C^{\infty}(\R^{2n})$.

{\bf BC Problem}\,. {\it Determine the trajectories $G(t|x_0,y_0)$ which begin on $\Lambda_0$ and
have the final position $x\in D_t = \Pi_1\Lambda_t$\,.\\
{\rm (i)} The Schr\"odinger case: Find the functions $x_0=x_0(t,x)$ and $y_0=y_0(t,x)$ from
\begin{eqnarray}\label{eq3.9}
    &\frac 12 \big[g(t|x_0 - \frac 12 Jy_0) + (x_0 + \frac 12 Jy_0)\big] = x\,,& \\
    \label{eq3.10}    &y_0=\grad \Phi_0(x_0)\,, \qquad x_0\in D_0\,.&
\end{eqnarray}\noindent
{\rm (ii)} The Heisenberg case: Find the functions $x_0=x_0(t,x)$ and $y_0=y_0(t,x)$ from}
\begin{eqnarray}\label{eq3.6}
    &\frac 12 \big[g(t|x_0 - \frac 12 Jy_0) + g(t|x_0 + \frac 12 Jy_0)\big] = x\,,& \\
    \label{eq3.7}    &y_0=\grad S_0(x_0)\,, \qquad x_0\in D_0\,.&
\end{eqnarray}

Lemma 2 justifies writing the left sides of (\ref{eq3.9}) and (\ref{eq3.6}) as the weighted sum
of the left and right flows. It is natural to interpret each of these equations as a midpoint
condition. Conditions (\ref{eq3.10}) and (\ref{eq3.7}) ensure that $(x_0,y_0)$ lies on
$\Lambda_0$. Observe that the initial density matrix (\ref{eq3.5}) is not real valued like
approximation (\ref{eq2.19}). However, by superposition with the complex conjugate of
(\ref{eq3.5}), the initial state can be made real.

    We treat the small time and big time solutions to the BC problem with different methods and
assumptions. The small time problem may be formulated as follows. Notice that the midpoint and
$\Lambda_0$ boundary conditions may be combined into a single statement, $\M_t(x') = x$.

\begin{defn}
\rm{(i)} In the Schr\"odinger problem set
\begin{equation}\label{eq3.11} \M_t(x')=\wt M_t(x') \equiv \sf
\big[g(t|x' - \sf J\grad \Phi_0(x')) + (x' + \sf J\grad \Phi_0(x'))\big] = x\,.\end{equation}
\vspace{-3 mm} \rm{(ii)} In the Heisenberg problem set
\begin{equation}\label{eq3.12} \M_t(x')=  M_t(x') \equiv \sf \big[g(t|x' - \sf J\grad S_0(x'))
 + g(t|x' + \sf J\grad S_0(x'))\big] =  x\,.
\end{equation}
\end{defn}

   The initial phase $\beta_0(x)$ in the extended Schr\"odinger problem is equal to either
$\Phi_0(x)$ or $S_0(x)$ depending on whether one has case (i) or (ii). Here it is assumed that
the support of $\beta_0(x)$ is all of $\R^{2n}$, and that $x$ is any point in $\R^{2n}$. If the
non-linear function $\M_t$ is invertible, then the BC problem has a single trajectory specified
by the initial values, $x_0(t,x) = {\M_t}^{-1}(x)$ and $y_0(t,x) = \grad\beta_0(x_0(t,x))$. The
arguments of the flows $g(t)$ are the left-right projections of $\Lambda_0$.

\begin{prop} \label{prop1} Let the Hessians $H''(x)$ and $\beta_0''(x)$ have $x$-uniform bounds
\begin{equation}
 \|H''(x)\|\leq c_1 < \infty,\qquad \|\beta_0''(x')\|\leq 2 \,,
\end{equation}
then there exists a $t_1>0$ such that the midpoint map $\M_t:\R^{2n}\rightarrow \R^{2n}$ is a
diffeomorphism for $t\in [-t_1,t_1]$. Specifically, $x_0(t,x) = \M_t^{-1}(x)$ is a unique
$C^1([-t_1,t_1],\R^{2n})$ solution of the BC problems {\rm(i)} and {\rm (ii)}, respectively.
\end{prop}

\noindent{\it Proof}. Consider the Heisenberg problem. For parameters $(t,x)\in
([-T,T],\R^{2n})$, define the family of maps $T_{t,x}:\R^{2n}\rightarrow \R^{2n}$,
\begin{equation*}\label{eq3.13}
    T_{t,x}(x') \equiv x + x' -M_t(x')\,.
\end{equation*}
The idea of the proof is to show that $T_{t,x}$ is a contraction mapping. Its unique fixed point,
$x^*(t,x)$, obeys $x=M_t(x^*(t,x))$  and establishes that $M_t$ is invertible. Connect an
arbitrary pair of points $(x_1,x_2)$ by the linear path $x(\xi) = x_1 + \xi(x_2 - x_1)$. Integrate
${dT_{t,x}(x(\xi))}/{d\xi}$ on $\xi\in[0,1]$ to show
\begin{equation*}\label{eq3.14}
    |T_{t,x}(x_2) - T_{t,x}(x_1)| \leq |x_2 - x_1|\int_0^1 \|\grad T_{t,x}(x(\xi))\| d\xi \,.
\end{equation*} Whenever the integral above is less than one for all $(x_1,x_2)$,
then $T_{t,x}$ is a contraction mapping.  Taking the $x'$ gradient of $T_{t,x}(x')$ it follows
that
\begin{eqnarray*}\label{eq3.15a}
    &\| \grad T_{t,x}(x')\| &\leq \frac 12 \Big[\|I - \grad g(t|l')\| + \|I - \grad g(t|r')\|
     + \frac 12 \|\grad g(t|l') - \grad g(t|r')\|
    \|S_0''(x')\|\Big] \vspace{2 mm} \\ \label{eq3.15b} &  &<2(\exp{\,c_1 |\,t|} -1) \,.
\end{eqnarray*}
To obtain the final inequality, $\|S_0''(x')\|$ was replaced by its bound 2, and the
Lemma~\ref{LemA3} (Appendix A) estimates of the Jacobi field $\grad g$ were employed. Let $t_1>0$
satisfy $2(\exp{(c_1 t_1)-1)}< 1$, then for times $|\,t|\leq t_1$ the rightmost combination above
is less than one and $T_{t,x}$ is a contraction mapping for $(t,x)\in [-t_1,t_1]\times \R^{2n}$.
An application of the implicit function theorem to $M_t(x_0(t,x))=x$ establishes that $x_0\in
C^1([-t_1,t_1],\R^{2n})$. The same argument applies to the Schr\"odinger BC problem. $\square$

The merit of the short time BC solution given in Proposition 1 is that it is global ---
applicable for all $(t,x)\in[-t_1,t_1]\times\R^{2n}$, and so in this time interval $\Lambda_t$
has a diffeomorphic $x$-projection.  Nevertheless, the results are restrictive. Typically only a
small time interval $[-t_1,t_1]$ is allowed, and the Hamiltonian and initial phase must have no
more than quadratic growth as $|\,x|\rightarrow \infty$.

The BC problem in the large time regime has a different geometry. Let $\Lambda_0$ have an
$x$-projection, $D_0$, that is contained within some compact set of $\R^{2n}$. As $\Lambda_t$
evolves away from $\Lambda_0$, it may develop into a multi-sheeted manifold.  Here, it is helpful
to recall some of the terminology related to caustic behavior of Lagrangian manifolds,
$\Lambda\subseteq\chi_2$.

\begin{defn} A point $m\in\Lambda$ is {\it non-singular} if it has a neighborhood that is diffeomorphically
projected onto $\R^{2n}_x$. If the opposite holds, it is {\it singular}. The set
$\Sigma(\Lambda)$, of all singular points, is called the {\it cycle of singularities}.  Its
complement is the {\it regular} set, $\RR= \Lambda / \Sigma(\Lambda)$.
\end{defn}

\begin{defn} The $x$ projection of the cycle of singularities is the {\it caustic  or focal set},
$\Pi_1(\Sigma(\Lambda)) = \K(\Lambda)$.  Let $D_t= \Pi_1 \Lambda_t$ denote the $x$-image of
$\Lambda_t$, and $\K(\Lambda_t)$ the associated caustic set. The {\it non-focal set} is defined
as the complement of the caustic set, $\N_t= D_t / \K(\Lambda_t)$.
\end{defn}

If $\Lambda_t$ has more than one sheet, the inverse of projection $\Pi_1:\Lambda_t \rightarrow
D_t$ ceases to be single valued: there are $x\in \N_t$ with multiple pre-images, $\{m^j\}_1^N
\subseteq \Lambda_t$, satisfying $\Pi_1 m^j = x$. The result below is stated for the Heisenberg
BC problem, with obvious adjustment it applies to the Schr\"odinger problem.

\begin{prop} \label{prop2} Let $\tilde x \in \N_t \subseteq D_t$ be a non-focal point with
pre-images $\{m^j \in \Lambda_t\}_1^N$. Then the Heisenberg BC problem has solutions
\begin{equation*}\label{eq3.16}
    M_t(x_0^j(t,x)) = x\,, \qquad j=1,\cdots, N\,.
\end{equation*}
The functions, $x_0^j(t,x)$,  are smooth and have support in a neighborhood, $\N_t \subseteq\wt
U_j \ni\tilde x$.
\end{prop}

\noindent{\it Proof}. Let $F_t(x',x) = M_t(x') - x$, where $F_t$ is smooth on the domain
$(x',x)\in D_0\times \N_t$. The implicit function theorem (IFT) solutions generated from
$F_t(x',x) = 0\,,$ will construct $x_0^j(t,x)$.  In order to apply the IFT one must have a
solution pair $(x_0,x)$, where $F_t(x_0, x)=0$ and $\det \grad_{x'}F_t(x_0,x)\neq 0$.  Consider
one pre-image $m^j$ of $\tilde x$. Each $m^j\in \Lambda_t$ has a 1-1 association with a point in
$D_0$, via $x_0 = \Pi_1 G(-t|m^j)$. The set $(x_0,\tilde x)$ is a solution pair of $F_t$.  It
remains to show that the determinant is non-zero. Since $\tilde x \in \N_t$, Lemma \ref{LemA4}
implies that $\det \grad M_t(x_0) \neq 0$. Furthermore $\grad_{x'} F_t(x',x) = \grad M_t(x')$
shows that $\det \grad_{x'}F_t(x_0,\tilde x)\neq 0$. The IFT then establishes the existence of
$x_0^j(t,x)$ defined on a neighborhood, $\wt U_j \ni\tilde x$. If $F_t\in C^{p}, p\geq 1$, then
$x_0^j\in C^{p}$. $\square$

Once $\Lambda_t$ has become multi-sheeted, typically the regular set $\RR_t$ is divided by the
cycle of singularities into disjoint subsets $\{\Lambda_t^j\}_1^N$. On $\Lambda_t^j\subset
\RR_t$, the restricted projection $\Pi_1^j\Lambda_t^j \equiv D_t^j \subset \N_t$ is
diffeomorphic. In this circumstance, $x_0^j(t,x)$ has the explicit representation
\begin{equation}\label{eq3.19}
x_0^j(t,x)= \Pi_1 G(-t|\big(\Pi_1^j \big)^{-1}x)\,, \quad x\in D_t^j\,.
\end{equation}
Identity (\ref{eq3.19}) determines $x_0^j(t,x)$ directly without requiring an IFT construction.

As is evident from $J(x,t) = \det\grad\M_t(x)$, \cf (\ref{eq3.2b}), the gradient structures in
the BC problem determine the WKB amplitudes as functions of the left and right Jacobi fields,
$\grad g(t|l')$ and $\grad g(t|r')$. It is of interest to find a more geometric representation of
these amplitudes.  This is achieved by constructing a suitable $\chi_2$ Poisson bracket related
to the BC problem.  First characterize the initial manifold $\Lambda_0$ by the (independent)
constraint functions
\begin{equation*}
\zeta(x',y') = y' - \grad \Phi_0(x')\,, \qquad \eta(x',y') = y' - \grad S_0(x')\,.
\end{equation*}
The conditions $\zeta(x',y') = 0$ and $\eta(x',y') = 0$ define $\Lambda_0$ in our two problems. A
second set of functions must necessarily carry information about the $\H$ dynamics. Select the
midpoint flow functions $x(t|x',y')$ for this role. Then one finds

\begin{lem} \label{lem3a} In terms of the $\Lambda_0$ constraint functions, $\zeta$ and $\eta$,
the $\H_1$ and $\H_2$ amplitudes have the $\chi_2$ Poisson bracket representation
\begin{eqnarray} \label{eq3.n1}
    \det \grad \wt M_t(x') & = & \det \{x(t),\zeta\}_2(x',\grad \Phi_0(x'))\,.   \\ \label{eq3.n2}
    \det \grad  M_t(x') & = & \det \{x(t),\eta\}_2(x',\grad S_0(x'))\,.
\end{eqnarray}
\end{lem}
\noindent {\it Proof}. Consider the $\H_2$ case, an elementary calculation shows that
\begin{eqnarray*}
    \{x(t)_\alpha, \eta_\beta\}_2(x',\grad S_0(x')) & = & \frac 12 \big[ \grad g(t|l')
    (I - \frac 12 J S''_0(x')) + \grad g(t|r')(I + \frac 12 J S''_0(x'))\big] \\
    & = & [\grad M_t(x')]_{\alpha\beta}\,.
\end{eqnarray*}
The determinant of this matrix is (\ref{eq3.n2}). The same type of argument gives (\ref{eq3.n1}).
$\square$

This type of bracket representation of the amplitudes is common \cite{Litj90,OH81} in the SPS
context. The components of constraint $\zeta$ (and $\eta$) are commutative with respect the
$\chi_2$ bracket. This involution property is a necessary and sufficient condition \cite{AW71}
for the manifold $\Lambda_0$ to be Lagrangian. In addition, the midpoint variables $x(t)$ are
commutative. This is a result of the fact that the functions $x(t)$ take their values on PPS
plane which is $\chi_2$ Lagrangian.

Our boundary condition analysis is concluded by identifying a $\chi_1$ loop structure that is
created by each BC solution. These loops are constructed from two types of $\chi_1$ line segments
--- chords and trajectories.  The quantity $C(x_1,x_2)$ denotes the chord (the $\R^{2n}$ geodesic) from $x_1$ to
$x_2$. Trajectory type line segments are specified by
\begin{eqnarray*}\label{eqn3.16a}
&T^{+}(x_1,x_2) & \equiv \{g(\tau|x_1); (\tau:0\rightarrow t) \} \,,\qquad x_2=g(t|x_1)\,, \\
\label{3.16b} &T^{-}(x_1,x_2) & \equiv \{g(\tau|x_1); (\tau:0\rightarrow -t) \} \,,\qquad
x_2=g(-t|x_1)\,.
\end{eqnarray*}
The $\pm$ labelling distinguishes between forward and backward evolution.

 Consider the Schr\"odinger problem with mid-point BC solution $x_0(x,t)$. The
$\Lambda_0$ manifold requirement (\ref{eq3.10}), when stated in the left-right variables, is
\begin{equation}\label{eq3.20}
 l_0(t,x) = x_0(t,x) - \sf \grad \Phi_0(x_0(t,x))\,,
\qquad r_0(t,x) = x_0(t,x) +  \sf \grad \Phi_0(x_0(t,x))\,.
\end{equation}
After time displacement $t$, the $\H_1$ flow moves $l_0=l_0(t,x)$ and $r_0=r_0(t,x)$ into the new
positions $l_t=l_t(t,x)=g(t|l_0)$ and $r_t=r_t(t,x)=r_0$ , \cf Lemma 2(i).

The pair of points $(l_t,r_0)$ define  the chord $C(l_t,r_0)$, having midpoint $x$. However,
$(l_t,r_0)$ are also the endpoints of the linked line segments, $C(r_0,l_0) + T^{+}(l_0,l_t)$.
Altogether, for each $x\in\N_t$, one has (see Fig. 1) a three sided oriented loop
\begin{equation*}\wt L(t,x)= C(r_0,l_0) + T^+(l_0,l_t) + C(l_t,r_0)\,.
\end{equation*}

{\small\input p1 \qquad\qquad \small\input p2 \vspace{3 mm}}

\subsection{Evolution Representations}

    Based on these solutions of the BC problem the WKB approximation for $U(t,x)$ takes the
    following form.

\begin{theor}\label{thm1}  Suppose  $x$ is a non-focal point associated with the
manifold $\Lambda_t$.  Let $\{\wt L^j(t,x)\}_1^N$ be the family of loops generated by the
Schr\"odinger BC solutions $\{x_0^j(t,x)\}_1^N, x\in \N_t$. Then the $O(\hbar)$ approximation of
$U(t,x)$ is
\begin{eqnarray}\label{eq3.22}
&U^{sc}(t,x) &= \,\sum_{j=1}^N N_j(t,x) \exp\Big\{\frac i{\hbar}\Phi_j(t,x) -\frac {i\pi}2
m_j(t,x)\Big\}\,,
\\ \label{eq3.23} &\Phi_j(t,x) &= \Phi_0(x_0^j(t,x)) + \,\oint_{\wt L^j(t,x)} p\cdot dq
-H(l_0^j(t,x))\, t\,,\\
\label{eq3.24}&N_j(t,x) &= \,|\det\grad \wt M_t(x_0^j(t,x))|^{-1/2}N_0(x_0^j(t,x))\,.
\end{eqnarray}
\end{theor}

Before proceeding to the Heisenberg case, it is useful to make a number of remarks specific to
Theorem 1. An important special case occurs in the problem where $U(0,x)=1$. Here $\Phi_0\equiv
0$ and $\Lambda_0 \subseteq \R_x^{2n}$.  In particular, $l_0=x_0=r_0$, so the loop $\wt L^j(t,x)$
simplifies into one chord $C(l_t,l_0)$ with midpoint $x$ and one dynamical curve $T^+(l_0,l_t)$
connecting the chord end points. If $\Lambda_t$ has one sheet, then $N=1$, and the theorem above
gives the representation obtained in \cite{Mar79}. However, it should be noted that Marinov did
not investigate the BC problem. Instead his approach was to construct the phase
$\Phi=\Phi_1(t,x)$ by solving the Hamilton-Jacobi (H--J) equation associated
\cite{Mar79,KN78,Int82} with the evolution problem (\ref{eq1.4}),
\begin{equation}
\label{eq3.25}
\partial_t \Phi(t,x)-H(x-\sf J\nabla\Phi(t,x))=0\,,
\end{equation}
with the Cauchy initial data $\Phi(0,x) = 0$. Because of the highly non-linear character of
$H(x)$ this H--J problem is much more difficult to solve than the BC problem (\ref{eq3.11})
treated above. It is straight forward to verify that $\Phi(t,x)$ given by (\ref{eq3.23}) is a
solution of (\ref{eq3.25}).

Marinov's work established the central role of the phase $\Phi(t,x)$ in the semiclassical
approximation of $U(t,x)$. However, knowledge of the H--J equation in the form (\ref{eq3.25}),
and the use of $\Phi(t,x)$ as the generator of a canonical transformation, \cf (\ref{eq2.15x}),
goes back to Poincar\'e \cite{HP99}.

Quantum operator identities induce corresponding Weyl symbol identities.  A simple example of
this is the unitarity property, $U(t)=U(-t)^{\dag}$.  Its Weyl symbol image is
$U(t,x)=U(-t,x)^*$, which in turn requires that $\Phi(t,x)= -\Phi(-t,x)$ and $N(t,x)=N(-t,x)$.

The generalization of Theorem 1 to non-static Hamiltonians $H(t,x)$ is achieved if the phase
$\Phi_j(t,x)$ is modified by the replacement
\begin{equation*}
H(l_0^j(t,x))t \rightsquigarrow \int_0^t H(g(\tau|l_0^j(t,x))\, d\tau\,.
\end{equation*}

Adjusting the notation above to cover the Heisenberg problem is easy. In this case $l_0(t,x)$ and
$r_0(t,x)$ are defined by (\ref{eq3.20}) with $S_0$ replacing $\Phi_0$. The right point $r_t$ is
no longer the fixed point $r_0$, but instead is  $r_t = r_t(t,x)= g(t|r_0(t,x))$. The curve
between $r_t$ and $r_0$ is $T^{-}(r_t,r_0)$. The resultant BC loop is the four segment structure
(see Fig. 2)
\begin{equation*}
W(t,x) \equiv C(r_0,l_0)+ T^+(l_0,l_t) + C(l_t,r_t) + T^{-}(r_t,r_0)\,.
\end{equation*}
The loop $W(t,x)$ is like the Schr\"odinger BC loop $\wt L(t,x)$, but has an additional side
$T^{-}(r_t,r_0)$ inserted at the point $r_0$. However, there is another simpler loop generated by
the BC solution. Denote by $\gamma(t|C(r_0,l_0)) \equiv \{g(t|x'): x' \in C(r_0,l_0)\}$. This
curve is the flow translated image of the chord $C(r_0,l_0)$.  In general it is no longer a
straight line but it has end points $(r_t,l_t)$, so
\begin{equation*}
L(t,x) \equiv C(l_t,r_t) + \gamma(t|C(r_0,l_0))\, .
\end{equation*}
defines a two segment closed curve, see Fig. 3.  We refer to $L(t,x)$ as the Heisenberg loop.

{\small\input p3 \qquad\qquad\quad \small\input p4 \vspace{3 mm}}

\begin{theor}\label{thm2}  Suppose  $x$ is a non-focal point associated with the
manifold $\Lambda_t$.  Let $\{L^j(t,x)\}_1^N$ be the family of Heisenberg loops generated by the
BC solutions $\{x_0^j(t,x)\}_1^N, x\in \N_t$. Then the $O(\hbar)$ approximation of $\rho(t,x)$ is
\begin{eqnarray}\label{eq3.26}
&\rho^{sc}(t,x) &= \,\sum_{j=1}^N \alpha_j(t,x) \exp\Big\{\frac i{\hbar}S_j(t,x) -\frac {i\pi}2
m_j(t,x)\Big\}\,,
\\ \label{eq3.27} &S_j(t,x) &= S_0(x^j_0(t,x)) + \,\oint_{L^j(t,x)} p\cdot dq \,,\\
\label{eq3.28}&\alpha_j(t,x) &= \,|\det\grad M_t(x_0^j(t,x))|^{-1/2}\alpha_0(x_0^j(t,x))\,.
\end{eqnarray}
\end{theor}

\noindent {\it Proof}.  Within the present setup, direct computation establishes both Theorems 1
and 2. Consider the Heisenberg case. The $d\tau$ integral contribution to the $\chi_2$ WKB phase
in (\ref{eq3.2a}) has both a $y\cdot \dot x$ and a $\H_2$ contribution. The first of these is
\begin{eqnarray}
&\int_0^t y(\tau)\cdot {\dot x}(\tau) d\tau& = \sf \int_0^t \big(\dot g(\tau|l_0)+ \dot
g(\tau|r_0)\big)\cdot J \big(g(\tau|l_0)- g(\tau|r_0)\big) d\tau \label{eq3.24a}\\
\label{eq3.25a} &   & = \oint_{W(t,x)} p\cdot dq \,.
\end{eqnarray}
The top equality is the result of Lemma 2(ii). The lower equality follows from an integration by
parts. For example
\begin{equation*}
\int_0^t \dot g(\tau|l_0)\cdot J g(\tau|l_0) d\tau  =  2\int_0^t p(\tau|l_0)\cdot \dot q(\tau|l_0)
d\tau + (q_0^l\cdot p_0^l -q_t^l\cdot p_t^l)\,,
\end{equation*}
where  $(q_0^l,p_0^l)=l_0$ and $(q_t^l,p_t^l)=l_t$. The $d\tau$ integral is that of $p\cdot dq$
over the curve $T^+(l_0,l_t)$. The sum of all the surface terms from (\ref{eq3.24a}) is just the
integral over the two chords, $C(r_0,l_0) + C(l_t,r_t)$. Thus one has that the time integral
(\ref{eq3.24a}) is the loop integral (\ref{eq3.25a})

The $\H_2\,d\tau$ term may be rewritten in loop form by employing the Poincar\'e--Cartan identity
in Lemma \ref{LemC1}.  Setting the initial curve to be $\gamma_0 = C(r_0,l_0)$, the identity
(\ref{eqC.1}) is
\begin{equation*}
(H(l_0)-H(r_0))\,t = \oint_{C(r_0,l_0) + T^+(l_0,l_t) + g(t|C(l_0,r_0)) + T^-(r_t,r_0)} p\cdot dq
\,.
\end{equation*}
Adding all the loop integrals gives (\ref{eq3.27}).

  The amplitude expressions result from $M_t(x_0^j(t,x)) =
x$, and the fact that the exponentiated integral in (\ref{eq3.2c}) vanishes. This latter is a
consequence of ${\rm{tr}}\, \grad_x\grad_y H\circ l = {\rm{tr}}\,H''(-\frac 12 J)=0$, etc.
$\square$

The Hamilton--Jacobi equation for the $\rho(t,x)$ problem is
\begin{equation*}
\label{eq3.29}
\partial_t S(t,x)-H(x-\sf J\nabla S(t,x)) + H(x+\sf J\nabla S(t,x))=0\,.
\end{equation*}
All the $S_j(t,x)$ functions, provided by (\ref{eq3.27}), are solutions of this Heisenberg H--J
equation. As $t$ nears $0$, $\Lambda_t$ will have one sheet. Its phase obeys, $S(0,x)=S_0(x)$.

Lemma \ref{LemA4} establishes that the amplitudes $N_j(t,x)$ and $\alpha_j(t,x)$ are finite for
all $x\in \N_t$. When $x\in \K(\Lambda_t)$ the amplitudes diverge and the representations of
Theorem 1 and 2 are not applicable at these caustic points. If $x$ is not in $D_t$, it is
classically forbidden. In this sector, $U(t,x)$ and $\rho(t,x)$ are $O(\hbar^{\infty})$.

The area representation of $S(t,x)$ is a consequence of Stokes' theorem,
\begin{equation}\label{eq3.30a}
S(t,x) = S_0(x_0(t,x)) + \int_{\sigma(L(t,x))} dq\wedge dp\,.
\end{equation}
Above, $\sigma(L(t,x))$ denotes any oriented surface with boundary $L(t,x)$. The specific form of
(\ref{eq3.30a}) resulted form choosing the $\gamma_0$ curve to be $C(r_0,l_0)$.  Other natural
choices for the curve connecting $r_0$ and $l_0$ do occur.  Consider the problem where
$\rho_0(x)$ is given by the projection operator $\hat \rho_0 = |\psi_0\rangle \langle \psi_0|$.
As was noted in Sec.~2 the initial phase function $S_0(x)$ is the loop $C(l_0,r_0) +
\gamma_\lambda(r_0,l_0)$ integral (\ref{eq2.20}). By choosing $\gamma_0 =
\gamma_\lambda(r_0,l_0)$ in the identity  (\ref{eqC.1}) one can incorporate the initial phase
into the final loop integral. So for this particular initial state, the entire Heisenberg phase
has the form of a symplectic area
\begin{equation*}\label{eq3.30b}
S(t,x) = \int_{\sigma(L'(t,x))} dq\wedge dp\,,
\end{equation*}
where $L'(t,x) = g(t|\gamma_\lambda(r_0,l_0)) + C(l_t,r_t)$\,.

The phases $\Phi(t,x)$ and $S(t,x)$ form representations of the $\chi_2$ Lagrangian manifolds
$\Lambda_t$, via $y=\grad \Phi(t,x)$ and $y=\grad S(t,x)$. In turn these phases define, \cf
(\ref{eq2.15x}), a canonical transformation between the dynamical left right coordinates
$l(t|l_0)$ and $r(t|r_0)$\,.  The $\Lambda_t$ phases also generate dynamical flow.  In the
Heisenberg case one has
\begin{eqnarray*}
    l(t|l_0) = g(t|l_0) = x - \sf J \grad S(t,x)\,,\qquad
    r(t|r_0) = g(t|r_0) = x + \sf J \grad S(t,x)\,.
\end{eqnarray*}

The presence of symmetries add structure and simplify these semiclassical expansions.  Two Weyl
symbol specific symmetries are: affine canonical covariance and $\hbar$-parity invariance.

Consider the first of these.  In the notation of Appendix A, let $V$ be the unitary operator
defined by the affine transformation  $A:\chi_1 \rightarrow \chi_1$. Let $U_V(t)$ and $\rho_V(t)$
be the symbols of the $V$-similarity transformed evolutions $\hatw U_{V}(t)\equiv V\hatw
U(t)V^{-1}$ and $\hat \rho_{V}(t)\equiv V\hat \rho(t)V^{-1}$. The semiclassical covariance
statements are
\begin{equation*}
U^{sc}_{V}(t) = U^{sc}(t)\circ A^{-1} \,,\qquad  \rho^{sc}_{V}(t) = \rho^{sc}(t)\circ A^{-1} \,.
\end{equation*}
These identities  follow from the exact covariance property, \eg $\rho_{V}(t) = \rho(t)\circ
A^{-1}$, and the fact that asymptotic expansions have unique coefficients.

The $\hbar$ parity property is specific to Heisenberg evolution.  Recall  that, \cf
(\ref{eq2.16}), the symbol $\rho_0(x;\hbar)$ for the initial state $\hat \rho_0 = |\psi_0\rangle
\langle \psi_0|$ is an even function of $\hbar$.  For most problems of physical interest this
$\hbar$-even property is maintained under time evolution. Specifically, one has

\begin{lem} \label{lem4aa} (Evolution $\hbar$-parity). Suppose the real valued $H(x;\hbar)$ and $\rho_0(x;\hbar)$
have even $\hbar$ parity. Then the solution of (\ref{eq1.5}), $\rho(t,x)$, is real and has even
$\hbar$ parity.  The associated semiclassical expansion has the improved convergence
\begin{eqnarray*}
    &\rho(t,x) = &\rho^{sc}(t,x)\,[1 + O(\hbar^2)]\,,  \\
    &\rho^{sc}(t,x) = &|\det\grad M_t(x_0(t,x))|^{-1/2}\alpha_0(x_0(t,x))\cos[S(t,x)/\hbar]\,.
\end{eqnarray*}
\end{lem}

\noindent{\it {Proof}}. The function $\rho(t,x)=\rho(t,x;\hbar)$ is defined as the unique
solution of (\ref{eq1.5}).  Since the Moyal bracket and $H(x;\hbar)$ are $\hbar$ even, the
equation of motion (\ref{eq1.5}) is invariant under $\hbar \rightarrow - \hbar$. Because initial
state the $\rho_0(x;\hbar)$ is even it follows that $\rho(t,x; \hbar)$ is $\hbar$ even. $\square$

 It is useful to review to what extent the use of the secondary phase approach is
critical in obtaining Theorems 1 and 2.  Marinov obtained the small time, $\Phi_0=0$, version of
Theorem 1 without using the embedding of $\chi_1$ into  $\chi_2$ .  In this approach (see
\cite{Int82} for related details) one starts with problem (\ref{eq1.4}), and assumes a WKB
representation for $U(t,x)\approx U^{sc}(t,x)$. After applying the definition of the $*$ product
to $H*U^{sc}(t)$ the Hamilton-Jacobi equation equation (\ref{eq3.25}) for $\Phi(t,x)$ is derived
from the $\hbar\rightarrow 0$ limit. The transport equations for amplitudes $N(t,x)$ are obtained
from (\ref{eq1.4}) by extracting the ordinary differential equations generated by the higher
order $\hbar$ term identities.  A similar approach would work for the $\rho(t,x)$ problem.  What
this strictly primary phase space approach misses is the recognition of the Lagrangian manifold
$\Lambda_t$ as the geometric structure that determines the Schr\"odinger and Heisenberg WKB
expansions.  Given the SPS setting one can make the WKB representations global by following
Maslov's approach.  In Theorems 1 and 2 the base Lagrangian plane was always taken to be
$\R_x^{2n}$ with commutative coordinates $\{x_\alpha\}_1^{2n}$.  The caustic and
non-diffeomorphic behavior of the $\Pi_1$ projection is avoided by replacing $\Pi_1$ by a
different and non-singular projection:  $\Pi^*\Lambda_t = \Lambda^*$. Here $\Lambda^*$ is a a new
Lagrangian plane with  mixed (but commuting) coordinates
$\{x_{\alpha(i)},y_{\beta(i)}:i=1,\dots,2n\}$.  A related point concerns the amplitude
representations. Only in the SPS setting is it possible to express the amplitudes as determinants
of a Poisson bracket.

%% file: p1.tex
$\lefteqn{\mbox{\unitlength=1pt
\begin{picture}(108.000 ,108.000 )
\put(10,0){Figure 1. The Schr\"odinger loop.}
\put( 120.000 , 40.000 ){$\R_q^n$}
\put( 20.000 , 125.000 ){$\R_p^n$}
\put( 50.000 , 35.000 ){$x_0$}
\put( 27.000 , 50.000 ){$l_0$}
\put( 100.000 , 70.000 ){$x$}
\put( 77.000 , 35.000 ){$r_0$}
\put( 123.000 , 115.000 ){$l_t$}
\put( 15.000, 101.000){$T^+(l_0,l_t)$}
\end{picture}}}
{\mbox{\includegraphics{pi1.ps}}}$

%% file: p2.tex
$\lefteqn{\mbox{\unitlength=1pt
\begin{picture}(108.000 ,108.000 )
\put(10,0){Figure 2\,. The Heisenberg BC solution.}
\put( 120.000 , 40.000 ){$\R_q^n$}
\put(20.000 , 125.000 ){$\R_p^n$}
\put( 20.000 , 50.000 ){$l_0$}
\put( 100.000 , 35.000 ){$r_0$}
\put(70.000, 110.000 ){$l_t$}
\put( 120.000 , 80.000 ){$r_t$}
\put( 90.000 , 95.000 ){$x$}
\put(60.000, 40.000 ){$x_0$}
\put( 105.000 , 55.000 ){$x_r$}
\put( 55.000 , 75.000 ){$x_l$}
\end{picture}}}
{\mbox{\includegraphics{pi2.ps}}}$

%% file: p3.tex
$\lefteqn{\mbox{\unitlength=1pt
\begin{picture}(108.000 ,108.000 )
\put(10,0){Figure 3. Heisenberg loop area $L(t,x)$.} \put( 120.000 , 40.000 ){$\R_q^n$} \put(
20.000 , 120.000 ){$\R_p^n$} \put( 58.000 , 36.000 ){$x_0$} \put( 28.000 , 39.000 ){$l_0$} \put(
95.000 , 119.000 ){$x$} \put( 77.000 , 44.000 ){$r_0$} \put( 68.000 , 127.000 ){$l_t$} \put(
121.000 , 129.000 ){$r_t$} \put( 86.000 , 82.000 ){$r_\tau$} \put( 35.000 , 72.000 ){$l_\tau$}
\end{picture}}}
{\mbox{\includegraphics{pi3.ps}}}$

%% file: p4.tex
$\lefteqn{\mbox{\unitlength=1pt
\begin{picture}(108.000 ,108.000 )
\put(10,0){Figure 4. $S(t,x)$ as an evolving front.} \put( 120.000 , 40.000 ){$\R_q^n$} \put(
20.000 , 120.000 ){$\R_p^n$} \put( 58.000 , 35.000 ){$x_0$} \put( 28.000 , 39.000 ){$l_0$} \put(
95.000 , 119.000 ){$x$} \put( 77.000 , 44.000 ){$r_0$} \put( 68.000 , 126.000 ){$l_t$} \put(
121.000 , 129.000 ){$r_t$} \put( 86.000 , 82.000 ){$r_\tau$} \put( 35.000 , 72.000 ){$l_\tau$}
\end{picture}}}
{\mbox{\includegraphics{pi4.ps}}}$

%% file: sec4.tex
\section{Applications and Examples}
\setcounter{equation}{0} In this section we collect various interpretations, extensions and
applications of Theorems 1 and 2. For example, in both the Schr\"odinger and Heisenberg pictures,
time evolution is implemented by a one parameter group.  Define $\Gamma(t)\hat \rho_0 \equiv
U(t)\hat \rho_0 U(t)^{\dag}$. These two group properties  are $U(t_1+t_2) = U(t_1)\,U(t_2)$ and
$\Gamma(t_1+t_2)=\Gamma(t_1)\circ\Gamma(t_2)$, respectively.  What are the symplectic phase
identities generated by these groups? Another question of interest is to establish the mutual
consistency of the $U^{sc}(t,x)$ and $\rho^{sc}(t,x)$ approximations.

The systems (\ref{eq1.4}) and (\ref{eq1.5}) admit exact solutions if the Hamiltonians are
quadratic. In this special case, the contrasting behavior of $U(t,x)$ and $\rho(t,x)$ is
explored. Finally, we show how the semi-classical solutions of (\ref{eq1.5}) behave when $
\rho_0(x)$ is not a rapidly oscillating function of $\hbar$.

\subsection{Phase Additivity Identities}

Phase additivity identities,  mirroring the group properties of $U(t)$ and $\Gamma(t)$, are a
direct consequence of the loop structure of the BC problem.  Denote the BC loop integrals by
\begin{equation*}
\varphi(t,x;\wt L) = \oint_{\wt L(t,x)} p\cdot dq\,, \qquad \A(t,x;W) = \oint_{W(t,x)} p\cdot
dq\,.
\end{equation*}

First consider the Schr\"odinger problem. Let $t_1$ and $t_2$ be successive time displacements.
For $x\in \N_{t_1+t_2}$ the BC solution determines, $x_0(t_1+t_2,x)$, (where the $j$ label is
suppressed). The loop for the full time interval is
\begin{equation*}
\wt L_{1+2}(t_1+t_2,x) \equiv T^+(l_0,l_{1+2}) + C(l_{1+2},r_0) + C(r_0,l_0)\,.
\end{equation*}  Here $l_{1+2}=g(t_1+t_2|l_0)$.  As Fig.~5 shows this loop can be decomposed into
the sum of two (or more) sub-loops defined for $t_1$ and $t_2$, \eg $l_1=g(t_1|l_0)$.  Let $\wt
L_1$ and $\wt L_2$ be the loops with endpoints $(l_0,l_{1},r_0)$ and $(l_1,l_{1+2},r_0)$. It
follows from their definitions that these loops are additive,
\begin{equation*}
\wt L_{1+2}(t_1+t_2,x) = \wt L_{1}(t_1,x_1) + \wt L_{2}(t_2,x)\,, \qquad x_1=\frac 12
(l_1+r_0)\,.
\end{equation*}
\vspace{-5 mm}

{\small\input p5\qquad\qquad\quad \small\input p6 \vspace{3 mm}}

  As Fig.~6 makes clear there is a corresponding loop additivity in the
Heisenberg BC problem. Note that there is no requirement that the intermediate midpoint
$(t_1,x_1)$ be non-focal. Using the phase definitions above one obtains the following.

\begin{prop} \label{prop3} Let $x\in \N_{t_1+t_2}$ and suppose, $\wt L^j_{1+2} = \wt L^j_{1}+\wt
L^j_{2}$; $W^j_{1+2} = W^j_{1}+W^j_{2}$, are loop decompositions consistent with the BC solutions
$\wt M_{t_1+t_2}(x_0^j(t_1+t_2,x))=x$ and $M_{t_1+t_2}(x_0^j(t_1+t_2,x))=x$, $j=1,\dots, N$, then
\vspace{-2 mm}
\begin{eqnarray}\label{eq4.2a} &\varphi(t_1+t_2,x;\wt L^j_{1+2})& =\,
\varphi(t_1,x_1^j;\wt L^j_{1})+ \varphi(t_2,x;\wt L^j_{2})\,,
\\ \label{eq4.2b}&\A(t_1+t_2,x; W^j_{1+2})& =\, \A(t_1,x_1^j; W^j_{1})+ \A(t_2,x;
W^j_{2}) \,.
\end{eqnarray}
\end{prop}

Consider the short time regime where Proposition 1 holds. Let $\Phi_0 = 0$ and $x_0(\tau,x) = \wt
M^{-1}_\tau(x)$ be the BC solution. In this sector, it is possible to compare the rule
(\ref{eq4.2a}) with the phase addition formula established by Marinov \cite{Mar79}. Denote by
$X(\tau,x)$ the two segment loop $T^+(x_0,g(\tau|x_0)) + C(g(\tau|x_0),x_0)$, where $x_0=l_0=\wt
M^{-1}_\tau(x)$. Suppose $\varphi(\tau,x;X)$ is the corresponding loop phase. In terms of these
two sided figures, see Fig.~7, the loop additivity is realized by
\begin{equation*}
X_{1+2}(t_1+t_2,x)= X_1(t_1,x_1) + X_2(t_2,x_2)+ C(l_0,l_1)+ C(l_1,l_2) + C(l_2,l_0)\,.
\end{equation*}
The implied phase additivity now reads
\begin{equation}\label{eq4.3}
\varphi(t_1+t_2,x;X_{1+2}) = \varphi(t_1,x_1;X_1)+ \varphi(t_2,x_2;X_2) + P_3(x,x_1,x_2)\,.
\end{equation}
Equation (\ref{eq4.3}) is the version of additivity found by Marinov. It is the use of the two
sided loops $X$, rather than the three sided loops $\wt L$, that accounts for the extra
triangular area $P_3$.


{\small\input p7 } \vspace{5 mm}

A somewhat similar additivity links the Schr\"odinger and Heisenberg phases with each other. On
the left branch of Fig.~2, $\varphi(t,x_l;X_l)$ is defined by a two sided loop,
$X_l(t,x_l)=T^+(l_0,l_t)+C(l_t,l_0)$. Similarly, $\varphi(t,x_r;X_r)$ is the right branch area
defined by $X_r(t,x_r)=T^+(r_0,r_t)+C(r_t,r_0)$. Recall from (\ref{eqnA4b}), that the four chord
loop with midpoints $(x,x_r,x_0,x_l)$ has symplectic area $P_4(x,x_r,x_0,x_l)$. With this
notation one has

\begin{prop} For $x\in \N_t$ let $(x,x^j_r,x^j_0,x^j_l)\,$ be the
midpoints of the Heisenberg BC loop solution $\{W^j(t,x)\}_1^N$. Then
\begin{equation}\label{eq3.31}
\A(t,x;W^j) = \varphi(t,x^j_l ;X_l^j) - \varphi(t,x^j_r;X_r^j) +
P_4(x,x_r^j,x_0^j,x_l^j)\,,\qquad j=1,\cdots,N\,.
\end{equation}
\end{prop}
\noindent{\it Proof.} This follows directly from the loop decomposition
\begin{eqnarray*}\label{eq3.32}
W(t,x)  &=& C(r_0,l_0)  + T^+(l_0,l_t)  + C(l_t,r_t)  + T^{-}(r_t,r_0)\,, \\
          &=& X(t_1,x_l) - X(t_2,x_r) + C(r_0,l_0) + C(l_0,l_t)  + C(l_t,r_t) + C(r_t,l_0)
           \,.\qquad\qquad \square
\end{eqnarray*}

\subsection{Mutual Consistency of $U^{sc}(t)$ and $\rho^{sc}(t)$}

One may build the Heisenberg evolution $\rho(t,x)$ from its Schr\"odinger  components, via
$\rho(t,x) = U(t)*\rho_0 * \overline{U(t)}$.  The goal of this subsection is to show that the use
of the approximation $U(t,x)\thickapprox U^{sc}(t,x)$ combined with the evaluation of the $*$
product by the stationary phase approximation constructs $\rho^{sc}(t,x)$.   This calculation
will be carried out using the single sheet BC solutions found in Proposition~1. All the
amplitudes and phases of $U^{sc}(t,x)$ and $\rho^{sc}(t,x)$ will be written without the
multi-sheet subscript, $j$. This type of composition argument was used by Berry and Balazs
\cite{BB79} to construct $\rho^{sc}(t,x)$ in the case where $\R^n_q$ is one dimensional.

First observe that the required form of the $U^{sc}(t,x)$ is the Marinov version obeying the
initial condition $U(t,x)=1$. Since $\Phi_0=0$, the amplitude takes the simplified form
\begin{equation}\label{eq4.5a}
N(t,x) = \big[\det \sf (1+\grad g(t|\wt M_t^{-1}(x)))\big]^{-1/2}\,.
\end{equation}
Employing the three function $*$ product (\ref{eqnA9}), the approximate $\rho(t,x)$ is
\begin{eqnarray}
\lefteqn{[U^{sc}(t)*\rho_0* \overline{U^{sc}(t)}\,](x) \equiv \I_3(t,x)} \nonumber \\ & & = c\int
U^{sc}(t,x_1)\rho(x_2)\overline{U^{sc}(t,x_3)}\,\delta(x_1-x_2+x_3-x)\,
e^{iP_3(x_1,x_2,x_3)/\hbar}\, dx_1dx_2dx_3\,,\qquad \nonumber \\ \label{eq4.5aa} & & = c\int
N(t,x_1)\,N(t,x_3)\,\alpha_0(x_2)\, e^{i[\varphi(t,x_1)-\varphi(t,x_3) + S_0(x_2) +
P_3(x_1,x_2,x_3)]/\hbar}\, dx_1dx_3 \,.
\end{eqnarray}
where, $x_2=x_1+x_3-x$ and $c=(\pi \hbar)^{-2n}$.

Denote the total phase in (\ref{eq4.5aa}) by
\begin{equation}\label{eq4.6}
\Theta(x_1,x_3;t,x) = \varphi(t,x_1) - \varphi(t,x_3) + S_0(x_1+x_3-x) + P_3(x_3,x,x_1)\,.
\end{equation}
Let $\xi=(x_1,x_3)$ and $\Theta''(\xi;t,x)$ be the Hessian of $\Theta$ with respect to the $\xi$
variables. If $x_1=x_1(t,x)$ and $x_3=x_3(t,x)$ is an isolated non-degenerate critical point of
$\Theta$, \ie the solution of  $\grad_\xi\Theta(\xi;t,x)=0$, then the standard \cite{MF81}
stationary phase approximation to (\ref{eq4.5aa}) states
\begin{eqnarray}
&\I_3^{stph}(t,x) = &2^{2n} N(t,x_1)N(t,x_3)\,\alpha_0(x_2)|\det \Theta''(x_1,x_3;t,x)|^{-1/2}
\nonumber
\\ \label{eq4.7} & &\,\, \times\exp{\left[\frac i{\hbar} \Theta(x_1,x_3;t,x)+\frac {i\pi}4
{\rm{sgn}} \Theta''(x_1,x_3;t,x)\right]\,.}
\end{eqnarray}
Above, the functional form of the constraint condition is $x_2(t,x)=x_1(t,x)+x_3(t,x)-x$.

    In component form, the critical point equation reads
\begin{eqnarray}\label{eq4.8a}
&x_1& =\,\, x+ \sf J\grad \varphi(t,x_3) - \sf J\grad S_0(x_1+x_3 -x)\,, \\
\label{eq4.8b} &x_3& =\,\, x+ \sf J\grad \varphi(t,x_1) + \sf J\grad S_0(x_1+x_3 -x)\,.
\end{eqnarray}
For small times and under the same assumptions as Proposition 1, one can show that the system
above defines a contraction mapping and so has unique solution $x_1=x_1(t,x)$, $x_3=x_3(t,x)$. In
fact the set of points $\{x_2(t,x),x_1(t,x),x,x_3(t,x)\}$ are exactly the four midpoints
$\{x_0,x_l,x,x_r\}$ appearing in Fig.~2. The constraint condition for $x_2$ permits one to
replace the $P_3(x_3,x,x_1)$ area by the $P_4(x_3,x,x_1,x_2)$, \cf (\ref{eqnA10}). Substituting
these critical point arguments into the total phase and using Proposition~4 one has
\begin{equation*}\label{eq4.9a}
\Theta(x_1(t,x),x_3(t,x);t,x) = S(t,x)\,.
\end{equation*}
 This demonstrates that the stationary phase produced by $U^{sc}(t)*\rho_0*
\overline{U^{sc}(t)}$ is the same as the Heisenberg WKB phase $S(t,x)$.

Verifying amplitude consistency is more elaborate. A non-trivial determinant identity for
$\Theta''$ in terms of the individual left-right flows is needed.  Let $K_l=\grad g(t|l),\,
l=x_2(t,x) - \frac 12 \grad JS_0(x_2(t,x))$ and $K_r=\grad g(t|r),\, r=x_2(t,x) + \frac 12 \grad
JS_0(x_2(t,x))$. Using identities obtained from (\ref{eq4.8a}) and (\ref{eq4.8b}) one can
establish
\begin{eqnarray}\label{eq4.9b1}
\lefteqn{\det\Theta''(x_1(t,x),x_3(t,x);t,x)\qquad\qquad} \nonumber \\ \label{eq4.9b2} & &=
2^{6n} \big (\det(1+K_l)(1+K_r)\big)^{-1} \det\big[K_l(1-\sf JS_0''(x_2)) + K_r(1+\sf
S_0''(x_2))\big]\,.\qquad\qquad
\end{eqnarray} Putting (\ref{eq4.9b2}) into (\ref{eq4.7}) recovers the Theorem~2 amplitude $\alpha(t,x)$.   In the small time sector all the determinants above are positive and
$sgn \Theta''(x_1,x_3;t,x) = 0$. Altogether one has the identity
\begin{equation}\label{4.9}
    [U^{sc}(t)*\rho_0 * \overline{U^{sc}(t)}]^{stph}(t,x) = \rho^{sc}(t,x)\,.
\end{equation}

\subsection{Quadratic Exactness}

If the Hamiltonian is a quadratic function, then the semiclassical quantities $U^{sc}(t,x)$ and
$\rho^{sc}(t,x)$ are exact solutions of (\ref{eq1.4}) and (\ref{eq1.5}).  Thus it is of interest
to obtain the explicit formulas for the phases and amplitudes which arise is this special case.

The general time dependent Weyl symbol Hamiltonian of quadratic form is
\begin{equation*}\label{eq4.10}
    H(t,x) = \frac 12\, x\cdot H''(t)\,x + H'(t)\cdot x\,,
\end{equation*}
where $H''(t)$ is a $2n\times 2n$ symmetric matrix and $H'(t)$ is a $2n$ component vector. Both
$H''(t)$ and $H'(t)$ are real and $t$-continuous. The quantum Hamiltonian, $\widehat H(t)$,
results if $x$ is replaced  by $\hat x$ in $H(t,x)$.

 The classical trajectories $g(t,s|x)$ generated by a quadratic $H(t,x)$ are  linear functions of $x$.
 Explicit representations of $g(t,s|x)$ are obtained as follows.

The dynamical equations for $g(t,s|x)$ and $\grad g(t,s|x)$ are the linear ODE's
\begin{equation}\label{eq4.11}
{d\over {dt}} g(t,s|x) = JH''(t)\,g(t,s|x) + JH'(t) \,,
\end{equation}\vspace{-3mm}
\begin{equation} \label{eq4.12}
\frac d{dt} \grad g(t,s|x) = JH''(t)\,\grad g(t,s|x) \,, \quad\quad\ \
\end{equation}
with initial conditions $g(s,s|x)=x$ and $\grad g(s,s|x)=I$. Both (\ref{eq4.11}) and
(\ref{eq4.12}) are Jacobi field equations, but (\ref{eq4.12}) is homogeneous, while
(\ref{eq4.11}) has an inhomogeneity, $JH'(t)\,$. Let $K(t,s)$ be the $Sp(2n)$ solution of
\begin{equation} \label{eq4.13}
\frac d{dt} K(t,s)  = JH''(t) K(t,s) \,, \quad\quad K(s,s)  = I\,.
\end{equation}
The solution of the system (\ref{eq4.13}) is unique, which implies that $\grad g(t,s|x) =
K(t,s)$.  In terms of $K(t,s)$, the forward $(t>s)$ and backward $(s<t)$ flows are
\begin{eqnarray}\label{eq4.14a}
& g(t,s|x)  &= K(t,s)[\,x + F(t,s)] \,, \\ \label{eq4.14b} & g(s,t|x)&= K(s,t)\,x - F(t,s) \,,\\
\label{eq4.14c} &F(t,s) &= \int_s^t d\tau JK(\tau,s)^{T} H'(\tau) \,\qquad t\geq s\,.
\end{eqnarray}
The solution (\ref{eq4.14a}) is verified by putting it into (\ref{eq4.11}) and using the
symplectic identity $K(t,s)JK(t,s)^T = J$. From (\ref{eq4.14a}) one obtains (\ref{eq4.14b}) by
using the composition laws $g(t,s)\circ g(s,t)=Id$ and $K(s,t)K(t,s)=I$.

The behavior of the Heisenberg evolution is summarized in the following lemma. Note that for time
dependent Hamiltonians the boundary condition problem needs to be stated in terms of $g(t,s)$. So
the left side of (\ref{eq3.12}) has two times, \eg $M_{t,s}(x')$.

\begin{lem}\label{Lem4.1} For quadratic Hamiltonians  the semiclassical density
matrix is the exact solution of (\ref{eq1.5}), $\rho^{sc}(t,x)=\rho(t,x)$. The structure of
$\rho^{sc}(t,x)$ is given by:

{\rm(i)}~The BC problem $M_{t,s}(x_0(t,x))=x$ has a unique global solution given by the backward
flow
\begin{equation}\label{eq4.15a} x_0(t,x) = K(s,t)\,x - F(t,s) \,,
  \end{equation}\vspace{-6mm}

{\rm(ii)} \begin{equation} \label{eq4.15b} \qquad  \,\,
 \quad S(t,x) = S_0(x_0(t,x))\,, \qquad \det \grad M_{t,s}(x') = 1\,, \end{equation}
\vspace{-6mm}

{\rm(iii)}  \begin{equation}\label{eq4.15c}  \qquad \quad \rho^{sc}(t,x) =
\rho_0(x_0(t,x))\,e^{iS_0(x_0(t,x))/\hbar} \,. \end{equation}
\end{lem}

\noindent {\it{Proof} }. The $x$-linearity of (\ref{eq4.14a}) combined with (\ref{eq3.12})
implies $M_{t,s}(x') = g(t,s|x')$.  Since $g(t,s)$ has inverse $g(s,t)$, (\ref{eq4.15a}) holds.
The phase identity in (\ref{eq4.15b}) follows from the fact that chords evolve into chords,
specifically $g(t,s|C(l_0,r_0)) = C(l_t,r_t)$. For this reason the Heisenberg loop $L(t,x)$
collapses to zero, so that all the WKB phase resides in the $S_0$ term.  The determinant identity
is a consequence of  $\grad M_{t,s}(x')$ being a symplectic matrix. That
$\rho^{sc}(t,x)=\rho(t,x)$ is verified by direct substitution into (\ref{eq1.5}). $\square$

The $x,y$ representation of the manifold $\Lambda_t$ is
\begin{equation*} y(t,x) = \grad S_0(K(s,t)\,x - F(t,s)) K(s,t)\,. \end{equation*}
This formula shows that $\Lambda_t$ is single sheeted.  Statement (ii) establishes that
$\Lambda_t$ is non-singular for all $t$.

Next consider Schr\"odinger evolution. Let $\sigma(s)\equiv \{t\in \R|\, \det(K(t,s) -I) = 0\}$
denote the set of caustic times where the amplitude function of $U^{sc}(t,x)$ diverges.  The
behavior of Schr\"odinger evolution is summarized below. These are known results
(\cite{Mar79,AH84,GB89}) and their proof parallels that of Lemma \ref{Lem4.1}.

\begin{lem}\label{Lem4.2}  For quadratic Hamiltonian Schr\"odinger systems with initial
data $U(0,x)=1$, the semiclassical solution of (\ref{eq1.4}) is exact at all non-caustic times.
The structure of $U^{sc}(t,x)=U(t,x)$ for $t \in \R /\sigma(s)$ is given by

{\rm(i)}~The BC problem $\wt M_{t,s}(x_0(t,x))=x$ has a  solution given by:
\begin{equation}\label{eq4.16a} x_0(t,x) = (K(t,s)+I)^{-1}(2x - F(t,s))    \,, \end{equation}\vspace{-6mm}

{\rm(ii)} \begin{equation} \label{eq4.16b} \qquad  \,\, N(t,x_0)= 2^n |\det[K(t,s)+I]|^{-1/2}\,,
 \qquad \end{equation}
\vspace{-6mm}

{\rm(iii)}  \begin{equation}\label{eq4.16c} \Phi(t,x) = - x\cdot J \frac {K(t,s)-I}{K(t,s)+I}\,x
-2J \frac {K(t,s)}{K(t,s)+I}\, F(t,s)\,x + C(t,s) \,,  \end{equation} where $C(t,s)$ is an
$x$-independent constant.
\end{lem}

The manifold $\Lambda_t$ is determined by the identity
\begin{equation*}
y(t,x) = -2J \frac {1}{K(t,s) + I} \big[ (K(t,s) - I)\,x + K(t,s)F(t,s) ]\,.
\end{equation*}
The $x$-linearity of $y(t,x)$ means that $\Lambda_t$ is a hyperplane. For each $t_c \in
\sigma(s)$, where $K(t_c,s)$ has a -1 eigenvalue, $\Lambda_{t_c}$ is singular and
$|\,y(t_c,x)|=\infty$.  As $t\rightarrow t_c$, $U^{sc}(t,x)$ has a delta function behavior.

In comparing these Heisenberg and Schr\"odinger solutions, it is evident that the main difference
is that $\rho(t,x)$ is always caustic free, while $U(t,x)$ has singular behavior at $t=t_c$.

\subsection{Evolution of semiclassically admissible observables}

An important special case of the Heisenberg evolution occurs when the initial operator $\hat
\rho_0$ is semi-classically admissible, and as a result $S_0=0$.  Most observables of physical
interest (energy, momentum, angular momentum and their products) are semi-classically admissible.

How does Theorem 2 simplify in this case?  First note that $S_0=0$ implies that the initial
Lagrangian manifold $\Lambda_0 = \R^{2n}$. Furthermore, the BC equation (\ref{eq3.14}) becomes
$M_t(x') = g(t|x) = x$, and so $M_t^{-1}(x) = g(-t|x)$ for all $t,x$.  From this it follows that
 $\det\grad M_t(x') = 1$ and so there are no caustics. Since $S_0 = 0$, the left, right and midpoint BC
solutions are the same $x_0(t,x)=l_0(t,x)=r_0(t,x)$. This means the left, right, and midpoint
flows collapse to a common trajectory and as a result the Heisenberg loop $L(t,x)$ reduces to a
point.  As a consequence $S(t,x)=0$.  Altogether one has
\begin{equation}\label{eq4.16}
\rho^{sc}(t,x) = \rho_0(g(-t|x))\,.
\end{equation}

This simple transport expression for $\rho^{sc}(t,x)$ is the point of departure
\cite{Pro83,DR87,BS91,OM95,McQ98} for higher order $\hbar$ expansions of $\rho(t,x)$.  Identity
(\ref{eq4.16}) is an example of Egorov's theorem \cite{Eg69}.

The fact that $S(t,x)$ remains zero is in marked contrast to the behavior of $U^{sc}(t,x)$ where
the phase $\Phi(t,x)$ spontaneously becomes non-zero even if its $t=0$ value vanishes. This
Heisenberg null phase stability is a consequence of the symmetry $\H_2(x,y)=-\H_2(x,-y)$. In
turn, this symmetry means that $y(t)=y_0$ is a constant of motion for $\chi_2$ trajectories that
have initial value $(x_0,y_0)=(x_0,0)$. Equivalently, $\Lambda_t=\Lambda_0$.


%% file: p5.tex
$\lefteqn{\mbox{\unitlength=1pt
\begin{picture}(108.000 ,108.000 )
\put(10,0){Figure 5. Schr\"odinger loop addition.}
\put( 120.000 , 40.000 ){$\R_q^n$}
\put( 20.000 , 120.000 ){$\R_p^n$}
\put( 50.000 , 35.000 ){$x_0$}
\put( 27.000 , 50.000 ){$l_0$}
\put( 100.000 , 70.000 ){$x$}
\put( 77.000 , 35.000 ){$r_0$}
\put( 123.000 , 115.000 ){$l_{1+2}$}
\put( 55.000 , 110.000 ){$l_1$}
\put( 72.000 , 74.000 ){$x_1$}
\put( 50.000 , 60.000 ){$\tilde L_1$}
\put( 90.000 , 100.000 ){$\tilde L_2$}
\end{picture}}}
{\mbox{\includegraphics{pi5.ps}}}$

%% file: p6.tex
$\lefteqn{\mbox{\unitlength=1pt
\begin{picture}(108.000 ,108.000 )
\put(10,0){Figure 6\,. Heisenberg loop addition.}
\put( 130.000 , 40.000 ){$\R_q^n$}
\put( 20.000 , 120.000 ){$\R_p^n$}
\put( 20.000 , 55.000 ){$l_0$} 
\put( 103.000 , 38.000 ){$r_0$}
\put( 80.000, 112.000 ){$l_{1+2}$} 
\put( 125.000 , 85.000 ){$r_{1+2}$}
\put( 105.000 , 102.000){$x$}
\put( 95.000 , 69.000 ){$r_1$} 
\put(64.000 , 40.000 ){$x_0$}
\put( 30.000 , 90.000 ){$l_1$} 
\put( 60.000 , 72.000 ){$x_1$}
\put( 70.000 , 58.000 ){$L_1$}
\put( 70.000 , 92.000 ){$L_2$}
\end{picture}}}
{\mbox{\includegraphics{pi6.ps}}}
$

%% file: p7.tex
$\lefteqn{\mbox{\unitlength=1pt
\begin{picture}(108.000 ,108.000 )
\put(10,0){Figure 7. Marinov loop addition.}
\put( 120.000 , 40.000 ){$\R_q^n$}
\put( 20.000 , 120.000 ){$\R_p^n$}
\put( 27.000 , 50.000 ){$l_0$}
\put( 80.000 , 72.000 ){$x$}
\put( 126.000 , 111.000 ){$l_{1+2}$}
\put( 60.000 , 116.000 ){$l_1$}
\put( 57.000 , 75.000 ){$x_1$}
\end{picture}
}}
{\mbox{\includegraphics{pi7.ps}}}
$

%% file: sec5.tex
\section{Conclusions}
\setcounter{equation}{0}

    The representations of operators by Weyl symbols gives one a complete statement of quantum
mechanics set in phase space. The Wigner transform is a Lie algebra isomorphism between operators
(with commutator bracket) and symbols (with Moyal bracket).  In this formalism, the Schr\"odinger
and Heisenberg evolutions, $U(t,x)$ and $\rho(t,x)$, are generated by Hamiltonians that are
pseudodifferential operators.  These two problems are made amenable to semiclassical
approximation by embedding  them in an extended phase space $\chi_2 = T^*\chi_1$. Set in
$\chi_2$, the two evolution problems become special cases of an extended Schr\"odinger evolution.

The geometry of the WKB approximation is controlled by $\Lambda_t$, the dynamical Lagrangian
manifold in $\chi_2$.  The associated BC problem is solved in terms of loops built from chords
and trajectories in PPS. The symplectic area of these loops form the WKB phase functions.  The
amplitudes are $\chi_2$ Poisson brackets constructed from the family of constraints that define
the initial and final BC Lagrangian manifolds, $\Lambda_0$ and $\N_t\subseteq\R^{2n}_x$.

An alternate characterization of Heisenberg evolution in symbol space is found in Marinov's
\cite{Mar91} path integral representation of $\rho(t,x)$.  This path integral has the following
structure. Write the mapping $\rho_0(x) \mapsto \rho(t,x)$ in terms of an integral transform
\begin{equation}\label{eq5.1}
    \rho(t,x) = \int dx'\,K(t;x,x')\,\rho_0(x')\,,
\end{equation}
where the kernel $K$ is $h^{-n}{\rm{Tr}}[U(t)\Delta(x')U(t)^{\dag}\Delta(x)]$. Define an action
functional by
\begin{equation*}
\S(t;\tilde x,\tilde y) = \int_0^t d\tau\, [\tilde y(\tau) {\cdot} \dot{\tilde{x}} (\tau)
 -\H_2(\tilde y(\tau),\tilde x(\tau)) ]\,.
\end{equation*} The path functions $\tilde x(\tau)$ obey the boundary conditions, $\tilde x(0)=x'$
and $\tilde x(t)=x$, while the functions $\tilde y(\tau)$ are unrestricted.  Then $K$ is the path
integral,
\begin{equation} \label{eq5.2}
K(t;x,x') = h^{-n} \int\!\!\! \int
D\tilde x\, D\tilde y\, e^{{i\S(t;\tilde x,\tilde y)}/\hbar} \,.
\end{equation} Here $D\tilde x\, D\tilde y$ are infinite products of
the dimensionless measures $h^{-n}dx$ and $h^{-n}dy$.

A second perspective of (\ref{eq5.2}) follows from the extended Schr\"odinger equation
(\ref{eq2.0}). In the SPS setting, $\rho(t,x)$  may be interpreted as a wave function generated
by the $\H_2$ evolution in $L^2(R^{2n}_x)$. The kernel $K$ is then the $\chi_1$ coordinate space
Dirac matrix element $\langle x|\exp\big({-i}\hatw\H_2 t /{\hbar}\big)|\,x'\rangle$ and
representation (\ref{eq5.2}) is the corresponding Feynman path integral. The parallels with the
$\rho^{sc}(t,x)$ approximation are evident. The functional $\S(t;\tilde x,\tilde y)$ has the same
form as the generic WKB phase (\ref{eq3.2a}), but with $\tilde y(\tau),\, \tilde x(\tau)$
functional arguments replacing the classical flow $G(t|\,x',y')$.

    The symplectic area WKB structure found for the evolutions $U(t)$ and $\rho(t)$ seems to be
universal. An alternate version of Theorem 1, suitable for quantum systems with external
electromagnetic fields, has been recently published in \cite{KO1}.  That paper introduced a gauge
invariant $*$ product of the Berezin type.  The $(x_1,x_2,x_3)$ triangle phase of this magnetic
$*$ product is a symplectic area with respect to the Faraday 2-form. The short time
electromagnetic generalization of our Theorem 1 is found in Theorem 6 of \cite{KO1}. Using the
SPS methods developed here combined with the with symbol calculus of the magnetic $*$ product,
the extension of Heisenberg WKB representation of Theorem 2 to a gauge invariant form is
straightforward.

Another direction of generalization is to consider non-Weyl $\hat q, \hat p$ quantum orderings.
The simplest of these alternate ordering schemes \cite{KN78} is
\begin{equation*}
\hat f(\hat q,\hat p) = f(\alpha \st{1}{\hat q} + (1-\alpha)\st{3}{\hat q}, \st{2}{\hat p})\,.
\end{equation*} where $\alpha\in [0,1]$. Normal, anti-normal, and Weyl ordering occur for
parameter values $\alpha = 0,1,\frac 12$. For all orderings with $\alpha \in (0,1)$ there will be
symplectic area WKB representation of Schr\"odinger and Heisenberg evolutions.

The extended phase space method developed here is expected to be a good platform for further
study. Since the $\chi_2$ Schr\"odinger evolution (\ref{eq2.0}) is unitary in $L^2(\R^{2n}_x)$,
norm error bounds for $\|\rho(t)-\rho^{sc}(t)\|_2$ may be obtained with standard
\cite{MF81,Int82} techniques. Higher order $\hbar$ expansions, with $U^{sc}(t,x)$ and
$\rho^{sc}(t,x)$ as first terms, are straightforward to derive using the known higher order
transport equations. Similarly the available globalization procedures \cite{MF81,KN78} readily
apply in the $\chi_2$ framework.

%% file: appendix1.tex
\section{The Star Product}
\setcounter{equation}{0}

 Comprehensive overviews of the Weyl symbol calculus are found in the books
\cite{KM91,Foll89}. In this appendix we summarize just the features of pseudodifferential
operators and the star product that our semiclassical analysis requires. It is useful to begin
with Weyl's original idea of quantization \cite{Wey27}. Let $\tilde f$ be the $\hbar$-Fourier
transform of $f:T^*\R^n\rightarrow\C$
\begin{equation}\label{eqnA1}
f(x) = \int_{\R^{2n}} dy\,\tilde f(y)\exp{\Big(\frac i{\hbar} y\cdot x\Big)} \equiv
\big(F_{\hbar,y\rightarrow x}\tilde f\big)(x)\,,
\end{equation}
then the Weyl quantization of $f$ is
\begin{equation}\label{eqnA2}
    \hatw f = \int_{\R^{2n}} dy\,\tilde f(y)\exp{\Big(\frac i{\hbar} y\cdot \hat x\Big)}
    \equiv f(\hat q,\hat p)\,.
\end{equation}
In this approach the family of unitary operators $\exp{\frac i{\hbar} y\cdot \hat x}$ is
associated with the exponential functions $\exp{\frac i{\hbar} y\cdot x}$. By Fourier
superposition this association is extended to construct a general operator $\hatw f$. The
notation $\hatw f = f(\hat q,\hat p)$ specifies a Weyl ordered function of the canonical
operators $(\hat q,\hat p)$.

Upon first inspection the quantization procedure above does not clarify if there is a well
defined inverse: given an arbitrary operator $\hatw f$ on $\H$ how is $f(x)$ determined? Moyal
solved this problem by showing that the inverse map $\hatw f \mapsto f$ is given by the Wigner
transform
\begin{equation}\label{eqnA3}
 f(x) =[\hatw f\,]\wig (x) \equiv \int dv e^{{-p\cdot v}/\hbar} \langle
q+{\scriptstyle{\frac 12}}v|\hatw{f}|q-{\scriptstyle{\frac 12}}v \rangle \, .
\end{equation}

Although there are a number of other consistent schemes \cite{BS91} which associate functions on
phase space with operators, normal ordering [$\hat p$ acts before $\hat q\, $], Wick ordering
[($\hat q + i \hat p\, $) acts before ($\hat q - i \hat p$)], etc., Weyl quantization has the
advantage that 1) it treats $\hat q$ and $\hat p$ symmetrically, 2) self-adjoint operators have
real symbols, and 3) the Moyal bracket is an even function of $\hbar$, in particular the
semiclassical correction to the leading Poisson bracket term is $O(\hbar^2)$, not $O(\hbar)$.

A basic symmetry of the Wigner--Weyl isomorphism is its covariance  with respect to affine
canonical transformations.  These transformations are parameterized by a matrix $R\in
Sp(2n)$ and a displacement $x_0$\,,
\begin{equation*}
x'= A(x) = R^{-1}(x-x_0)\,,\qquad A^{-1}(x') = R\,x' + x_0'\,.
\end{equation*}
The pair $(R,x_0)$ also defines a family of unitary operators, $V=V(R,x_0)$ obeying
\begin{equation*}
V\hat x V^{-1}  = A(\hat x) = R^{-1} \,(\hat x -x_0\,I)\,.
\end{equation*}
If $T(x_0)$ is the Heisenberg translation operator ($T(x_0)^{-1}\hat x T(x_0) = \hat x + x_0 I$) and $M(R)$ is a metaplectic operator (\ie a unitary
operator obeying $M^{-1} \hat x M = R\, \hat x$) then $V=T(x_0)M(R)$.  Affine covariance relates
the Weyl symbol of a $V$ similarity transformed operator to its $A$ transformed symbol.
Using the density matrix as an example, this universal covariance property \cite{OM95} is
\begin{equation}\label{eqnA3b}
    \big(V\hat \rho(t) V^{-1}\big)\wig \equiv \big(\hat \rho(t)_{V} \big)\wig =
     \hat \rho(t)\wig \circ A^{-1}\,.
\end{equation}

In classical mechanics, an observable $f(x)$ is a real valued function and the product of
observables is the commutative product of functions $f_1(x)f_2(x)$. In the Weyl symbol version of
quantum mechanics, the commutative product becomes the non-commutative star product. There are
three distinct types of representation of this star product; 1) the integral definition, 2) the
left-right operator characterization, and 3) Groenewold's exponential derivative formula
\cite{OM95,Gro46}. Each of these three forms is useful in a different way. The integral form is
convenient for proving the associativity of the $*$ product and determining the form of multiple
products; the left-right forms represent the $*$ composition as a pseudo-differential operator
(Lemma \ref{LemA2}), and the Groenewold formula gives explicit coefficients \cite{OM95} for the
small $\hbar$-asymptotic expansion of the star product.

Consider first the integral representation.  Let $\hatw f_j, j=1,\dots,N$ be $N$ different
operators. The Weyl symbol of their product is denoted by
\begin{equation*}\label{eqnA4}
    [\hatw f_1 \hatw f_2 \cdots \hatw f_N]\wig = f_1 * f_2 * \cdots * f_N \,.
\end{equation*}
The higher order $*$ product requires the following two functions
\begin{gather}\label{eqnA4b}
P_N(x_1,x_2,\dots,x_N) =2\sum_{i=1}^{N-1} \sum_{j>i}^N (-1)^{i+j+1} x_i\wedge x_j\,,
\qquad x_i\wedge x_j = x_i\cdot Jx_j\,, \\
S_N(x_1,x_2,\dots,x_N) = x_1-x_2 + \cdots + (-1)^{N+1}x_N\,.
\end{gather}
\begin{lem}\label{LemA1}
 Let $f_j\in L^1 (\R^{2n}) $ be Weyl symbols, $j=1,\dots, N$.  For even $N\geq 2$, the
integral form of the $*$ product is \begin{equation} \label{eqnA8} (f_1*f_2*\cdots *f_N)(x_{N+1})=
c^{N/2} \int d x_1\dots dx_N f_1(x_1)\cdots  f_N(x_N) \exp \big[ iP_{N+1}(x_1,\dots,
x_{N+1})/\hbar\big],
\end{equation} where $c=(\pi \hbar)^{-2n}$. For odd $N\geq 3$,
\begin{eqnarray}
\label{eqnA9}
{(f_1*f_2*\cdots *f_N)(x_{N+1})} &= &c^{(N-1)/2} \int d x_1\dots dx_N f_1(x_1)\cdots  f_n(x_N)
\qquad\qquad\qquad\qquad \nonumber\\
& &\delta\left(S_{N+1}(x_1,\dots,x_{N+1})\right) \exp \big[ iP_{N}(x_1,\dots, x_N)/\hbar\big].
\end{eqnarray}
\end{lem}

\noindent{\it Proof.} The result for $N=2$ follows directly \cite{OM95}  from (\ref{eqnA1}) and
(\ref{eqnA2}). The higher order products follow from the recurrence relation
\begin{equation} \label{eqnA10}
P_N(x_1,\dots,x_N) = P_{N-1}(x_1,\dots,x_{N-1})+(-1)^{N+1} 2x_N\wedge S_{N-1}(x_1,\dots,x_{N-1})\,.
\end{equation}
and an induction argument. $\quad \square$

The phase $P_N$ in the $*$ product has a simple geometric meaning. It is the action line-integral
\begin{equation}\label{eqnA11}
    P_N(x_1,\dots,x_N) = \oint_{L_N}p\cdot dq = \int_{\Sigma_N} dq\wedge dp\,.
\end{equation}
The geometric figure in the integral is an $N$-sided polygon in $T^*\R^n$ having
$(x_1,\dots,x_N)$ as the midpoints of its successive sides. Note that in Lemma \ref{LemA1} the
phase $P_N$ is needed only for $N$ odd. When $N$ is odd, this polygon is uniquely determined by
the midpoints $(x_1,\dots,x_N)$. Denote the boundary of this polygon by $L_N$ with orientation
$x_N\rightarrow x_{N-1}\dots \rightarrow x_1\rightarrow x_N$; $\Sigma_N$ is any oriented surface
with outer boundary $L_N$. Thus $P_N$ is the symplectic area of this polygon. For a single star
product $f_1*f_2$ the polygon is a triangle and $P_3$ is its symplectic area.

A variety of symmetries for the symplectic area, $P_N$, follow from its definition
(\ref{eqnA4b}). \vspace{-0 mm}
\begin{eqnarray}\label{1a}
P_N(x_1+a,\dots, x_N+a) & = & P_N(x_1,\dots,x_N)-\delta_{N}2a\wedge S_N(x_1,\dots,x_N)\,,\\
\label{2a} P_N(-x_1,-x_2,\dots, -x_N) & = & P_N(x_1,x_2,\dots, x_N)\,, \\
\label{3a} P_N(x_2,x_3,\dots, x_N, x_1) & = & P_N(x_1,x_2,\dots, x_N)\,, \\
\label{3b} P_N(x_N,x_{N-1},\dots, x_1) & = & -P_N(x_1,x_2,\dots, x_N)\,.
\end{eqnarray}
The above identities express translation, reflection, cyclic and anti-cyclic permutation
invariance. The quantity $\delta_{N}$ is 1 if $N$ is even and zero otherwise.

These geometrical results for $N=3$ are due to Berezin \cite{FAB72}. It is now known \cite{AW94}
that the $*$ product on all symmetric symplectic spaces  has a phase that is a triangle related
symplectic area. Other representations of the $N$-order $*$ product are found in references
\cite{OM95,OdA98}

 A second perspective on the $*$ product is to understand it as a $\Psi$DO.
These operators are generated by smooth SPS functions, which are also called symbols. A $\Psi$DO
acting on the function $f(x)$ in denoted by  $\H(\st{2}\X,\st{1}\Y)f(x)$. This operator is normal
ordered, which means that $\Y = -i\hbar \grad_x $ acts first followed by $\X$, which is
multiplication by $x$. The operator $\H(\st{2}\X,\st{1}\Y)$ is formally defined as follows,
\begin{equation}\label{ADO}
\big[\H(\st{2}\X, \st{1}\Y ) f\big](x) = \left[ F_{\hbar,y\rightarrow x} \H(x,y)
F_{\hbar,x\rightarrow y} f\right](x)\,.
\end{equation}
The presence of $\hbar$ in the Fourier transform $F_{\hbar,x\rightarrow y}$ means that
$\H(\st{2}\X, \st{1}\Y )$ is a function of $\hbar$, and that the $\hbar\rightarrow 0$ asymptotics
is easy to study.  Passing from a normal ordered symbol to a Weyl ordered symbol is easy and is
achieved by the transformation
\begin{equation}\label{ATR}
\H_{Weyl}(x,y) = \exp(\frac \hbar{2i}  \grad_x \cdot \grad_y)\H(x,y)\,.
\end{equation}
In this paper the symbols with SPS arguments $(x,y)$ generally correspond to normal ordered
operators, while symbols with the PPS argument $x$ correspond to Weyl ordering.

The definition of $\H(\st{2}\X, \st{1}\Y )$ is completed by placing $\H$ in a suitable class of
functions. Two important function classes enter in Maslov's proof of (\ref{eq3.1}). The first is
$T^m(\R^{4n}_z)$. This class denotes those $C^{\infty}(\R^{4n}_z)$ functions $\H$ which satisfy
the uniform growth estimates
\begin{equation*}\label{e1}
    |\grad_x^\gamma \grad_y^\rho \H(x,y)| \leq C_{\gamma\rho} (1+|x|)^m (1+|y|)^m\,,
     \quad |\gamma|,|\rho| \geq 0\,, \quad (x,y) \in \R_x^{2n}\times\R_y^{2n}=\R_z^{4n}\,,
\end{equation*}
where integer $m\geq 0$ and the positive constants $C_{\gamma\rho}$ are independent of $(x,y)$.
The symbols in the class $T^m(\R^{4n}_z)$ are $\hbar$ independent. Now widen this class so as to
include semiclassically admissible operators. These operators depend on $\hbar$ and admit an
asymptotic expansion about  $\hbar = 0$. One says $\H(x,y;\hbar)$ belongs to the class
$T^m_{+}(\R_z^{4n})$, if $\H\in C^{\infty}(\R^{4n}_z\times \R^{+})$ and has the $\hbar$ expansion
\vspace{-2mm}
\begin{equation}\label{eqnA11a}\vspace{-2mm}
    \H(x,y;\hbar) = \sum_{j=0}^J (i\hbar)^j \H_j(x,y) +
    \hbar^{J+1}\RR_J(x,y;\hbar)\,.
\end{equation}
The coefficient functions $\H_j\in T^m$, and the remainder term $\RR_J$ has the $(x,y;\hbar)$
uniform bound
\begin{equation*}
|\grad_x^\gamma \grad_y^\rho (\frac{\p}{\p \hbar})^{\lambda}\RR_J (x,y;\hbar)| \leq
C_{\gamma\rho\lambda} (1+|x|)^m (1+|y|)^m\,,
     \quad |\gamma|,|\rho|,\lambda \geq 0\,.
\end{equation*}

 Now return to the $*$ product.  Its left-right
multiplication forms are the following.\vspace{-1mm}
\begin{lem}\label{LemA2}  For all $f_1,f_2 \in S^{\infty} (\R^{2n})$\vspace{-1mm}
\begin{eqnarray}\label{eqnA12}
 f_1*f_2(x)& = & f_1(\st{2}\X -{\sf}J\st{1}\Y) f_2(x)\,,\\ \label{eqnA13}
 f_1*f_2(x)& =& f_2(\st{2}\X +{\sf}J\st{1}\Y) f_1(x)\,.
\end{eqnarray}
\end{lem}
\noindent{\it  Proof.} 
 Make a change of variables in (\ref{eqnA8}) so that
\begin{equation*}\label{eqnA14}
 f_1*f_2(x) = (2\pi\hbar)^{-2n} \int dx_2\,dy_2\, f_1(x
-\sf Jy_2)\,f_2(x_2)\exp\left({{i\over \hbar}y_2\cdot(x-x_2)}\right)\,.
\end{equation*}
In the integral one can replace $f_1(x-{1/2}Jy_2)$ with $f_1(\st{2}\X-{1/2}J \st{1}\Y)$.  Moving
this last factor outside the integral gives (\ref{eqnA12}).  A similar argument gives
(\ref{eqnA13}).$\quad \square$

Further review of the left right operator method in quantization and the use of the left right
projections to provide representations of SPS is found in \cite{KM91}.

\section{Jacobi Field Applications}
\setcounter{equation}{0}

The Jacobi field of a classical system characterizes its stability.  Given a trajectory $g(t|x)$,
the $2n \times 2n$ matrix solution of the linear system
\begin{equation} \label{eqB.1}
\frac {d}{dt} \grad g(t|x) - JH''(g(t|x))\grad g(t|x) = 0\,, \end{equation} \vspace{-6 mm}
\begin{equation}\label{eqB.1b}
 \grad g(0|x) = I\,,
\end{equation}
defines  the associated Jacobi field. The solutions of (\ref{eqB.1}, \ref{eqB.1b}) are symplectic
matrices.  For small times $\grad g(t|x)$ will not differ greatly from its initial value $I$.
This is made explicit in the following estimates. The norm $\|\bf{\cdot}\|$ denotes the operator
norm on $\R^{2n}$.

\begin{lem}\label{LemA3} Let the Hessian of $H$ have the $x$-uniform bound $\|H''(x)\|\leq c_1 <
\infty,$ then the Jacobi field has the growth estimates
\begin{equation}\label{eqB.2}\|\grad g(t|x)\| \leq e^{\,c_1 t}\,, \qquad
    \|\grad g(t|x) - I\| \leq (e^{\,c_1 t} - 1)\,.
\end{equation}
\end{lem}
\noindent{\it Proof}. Define $\xi(t|x) = \|\grad g(t|x)\|$ and use $\frac {d}{dt}\|\grad
g(t|x)\|\leq \|\frac {d}{dt} \grad g(t|x)\|$. Thus equation (\ref{eqB.1}) and the Hessian bound
implies
\begin{equation*}\label{eqB.3}
\frac {d}{dt}\xi(t|x) \leq \| JH''(g(t|x))\| \,\xi(t|x) \leq c_1\, \xi(t|x)\,,\qquad
\end{equation*}\vspace{-6 mm}
\begin{equation*}\label{eqB.4}
 \xi(t|x) \leq e^{\,c_1 t}\,.
\end{equation*}
Now estimate $\|\grad g(t|x) - I\|$. The integral equation equivalent of
(\ref{eqB.1},\,\ref{eqB.1b}) is
\begin{equation*}\label{eqB.5}
\grad g(t|x) = I + \int_0^t JH''(g(\tau|x)) \grad g(\tau|x)\,d\tau\,.
\end{equation*} Taking the norm of this relation gives
\begin{equation*}\label{eqB.6}
\|\grad g(t|x) - I\| \leq  \int_0^t \|JH''(g(\tau|x))\|\, \xi(\tau|x)\,d\tau
 \leq c_1 \int_0^t e^{\,c_1 t} \,d\tau\ =(e^{\,c_1t} - 1)\,.\quad\square
\end{equation*}

    Next we consider the detailed linkage between the BC problem, finite WKB amplitudes,
and the non-singular set of $\Lambda_t$ with respect to the projection $\Pi_1$. Although this
connection is implicit in the Maslov--Fedoriuk book \cite{MF81}, its precise statement is
important for the BC solvability analysis and the interpretation of the WKB amplitude
singularities.

First let us collect several definitions and facts.  Recall that $D_0$ is the support of $S_0$ or
$\Phi_0$ . Furthermore the projection $\Pi_1:\Lambda_0 \rightarrow D_0$ is diffeomorphic, with
explicit inverse $\Pi_1^{-1}(x') = (x',\grad S_0(x'))\,,\, x' \in D_0$. The regular subset of
$\Lambda_t =G(t)\Lambda_0$ is $\RR_t=\Lambda_t/ \Sigma(\Lambda_t)$. Let $T_m(\Lambda_t)$ denote
the tangent plane at $m\in \Lambda_t$. The statement, $m\in \RR_t$, means that no vector in
$T_m(\Lambda_t)$ has a zero $\Pi_1$ projection, \ie dim $\Pi_1 T_m(\Lambda_t) = 2n$. Since
$\Lambda_t$ is a Lagrangian manifold, $T_m(\Lambda_t)$ is a $2n$ dimensional symplectic vector
space for each $m$.  With this notation we have

\begin{lem}\label{LemA4}Let $x' \in D_0=\Pi_1\Lambda_0$\,, then
\begin{equation}\label{eqB.7} \det \grad M_t(x') \neq 0 \quad \Leftrightarrow \quad G(t|x',\grad S_0(x'))
\in \RR_t\,.
\end{equation}
\end{lem}

\noindent {\it Proof}.\  The $\chi_2$ Jacobi field $\grad G(t|m')$ maps vectors on
$T_{m'}(\Lambda_0)$ into vectors on $T_{G(t|m')}(\Lambda_t)$. The matrix $\grad G(t|m')$ is
symplectic, specifically $\det \grad G(\tau|m') =1$. Let $\{e_i\}_1^{2n}$ be a basis of $D_0
\subseteq\R^{2n}$ with origin at $x'$. Then $\{W_i=(e_i, S''_0(x')e_i)\}_1^{2n}$ is a basis of
$T_{m'}(\Lambda_0),\, m'=(x',\grad S_0(x'))$. Since $\grad G(t|m')$ is invertible $\{\grad
G(t|m')W_i\}_1^{2n}$ is a basis of $T_{G(t|m')}(\Lambda_t)$. The Jacobi field for $\H_2$ flow is
\begin{equation*}\label{eqB.8}\grad G(\tau|x',y') = \left[\begin{array}{cc} \frac 12 [\grad g(t|l') +
\grad g(t|r')\,] & \frac 14 [-\grad g(t|l') +\grad g(t|r')\,]J \vspace {3 mm} \\  J[\grad g(t|l')
-\grad g(t|r')\,] & -\frac 12 J[\grad g(t|l') +\grad g(t|r')\,]J
\end{array}\right]\,.\quad
\end{equation*}

Consider the conditions which ensure that $m\in \RR_t$. Let $Z=(Z_x,Z_y)$ represent an arbitrary
non-zero vector in $T_m(\Lambda_t)$. Since $\{\grad G(\tau|m')W_i\}_1^{2n}$ is a basis one has
\begin{equation*}\label{eqB.9}
Z = \sum_{i=1}^{2n} \lambda_i \grad G(\tau|m')W_i = \sum_{i=1}^{2n}\lambda_i \grad
G(\tau|m')\left( {\begin{array}{c}  e_i \\ S_0''(x')e_i  \end{array}}\right )\,,
\end{equation*}
where at least one of the constants $\lambda_i$ are non-zero. Project $Z$ onto $\R_x$, giving
\begin{equation*}\label{eqB.10a}
\Pi_1 Z = Z_x = \sf \frac {\p}{\p x'} \big[g(t|x'- \sf J\grad S_0(x'))
 + g(t|x'+\sf J\grad S_0(x'))\big] V\,,
\end{equation*} \vspace {-5 mm}
\begin{equation*}
    \label{eqB.10b}  \Pi_1 Z = \grad M_t(x')V\,, \qquad V = \sum_{i=1}^{2n}
  \lambda_i e_i \neq 0\,.
\end{equation*}
Thus a vector in $T_m(\Lambda_t)$ can have zero $\Pi_1$ projection iff $\det \grad M_t(x') = 0$.
$\square$

With the replacement of $S_0$ by $\Phi_0$ and $M_t$ by $\wt M_t$, the equivalence (\ref{eqB.7})
holds for the Schr\"odinger problem.

\section{A Poincar\'e--Cartan Identity}
\setcounter{equation}{0}

The following identity is required in the proof of Theorem 2.

\begin{lem}\label{LemC1}
Let $\gamma_0$ be an arbitrary smooth curve from $x_1$ to $x_2$. Let $\sigma(t|\gamma_0)$ be the
surface defined by the images of $\gamma_0$ under the flow $g(\tau,t_0)\,,\,\, t_0 \leq \tau \leq
t$. Then
\begin{equation} \label{eqC.1}
 \int_{\sigma(t|\gamma_0)} dq\wedge dp = \oint_{\p\sigma(t|\gamma_0)} p\cdot dq = \int_{t_0}^t \big [ H(\tau, g(\tau,t_0|x_1)) -
H(\tau, g(\tau,t_0|x_2)) \big ] d\tau\,.
\end{equation}
\end{lem}

{\it Proof sketch}. Decompose the curve $\gamma_0$ into segments that are singularity free with
respect to some Lagrangian plane.  For notational convenience denote this plane by $\R^n_q$.  Now
locally embed this curve into a non-singular Lagrangian manifold $\lambda_0$ having phase
$s_0(q)$. Denote by $\lambda_t = g(t,t_0|\lambda_0)$ the time evolution of $\lambda_0$. Extend
the original phase space $\chi_1$ by adding time to the $q$-coordinate manifold, $(q,t;p,p_t)\in
\chi_1^{+} = T^*R^{n+1}$.  Let $s(t,q)$ be the phase of $\lambda_t$, then the manifold
\begin{equation*}
\lambda^{+} \equiv \{(q,t;p,p_t)\in \chi_1^{+}|(p,p_t)=(\grad s(t,q), {\p_t} s(t,q))\}
\end{equation*}
is Lagrangian.  On $\lambda^{+}$ the 1-form $\omega^+ = p\,dq - H(t,q,p)\,dt$ is closed by virtue
of the H--J equation.  Closed loop integrals on $\lambda^+$ vanish and can be arranged to give
(\ref{eqC.1}). $\square$\break

{\bf Acknowledgments}.  The authors are indebted to M. V. Karasev for suggesting a number of
improvements to the paper. We are also grateful to our colleagues Steve Fulling and Frank Molzahn
for critically reading the manuscript. The research of T.A.O. is supported by a grant from
Natural Sciences and Engineering Research Council of Canada.